\documentclass[conference]{IEEEtran}

\usepackage[utf8]{inputenc} 
\usepackage[T1]{fontenc}
\usepackage{url}              % provides \url{...}
\usepackage{cite}     
\usepackage{placeins}% improves presentation of citations

\usepackage[cmex10]{amsmath}  % Use the [cmex10] option to ensure complicance
\usepackage{mleftright}       % fix to wrong spacing of \left-,
\mleftright                   % \middle- \right-commands 

\usepackage{graphicx}         % provides \includegraphics{...} to
\usepackage{booktabs}         % fixes poor spacing in tables and
\usepackage{algpseudocode,tabularx,diagbox}

\usepackage[font=footnotesize,labelsep=period]{caption}
\usepackage{subcaption}  

\usepackage{etoolbox}
\usepackage{amsfonts}
\usepackage{amsmath}
\usepackage{xcolor}
\usepackage{bbm}
\usepackage{acronym}
\usepackage{amssymb}
\usepackage{comment}
\usepackage{paralist}
\usepackage{makecell}
\usepackage{tikz}
\usepackage{pgf,pgfplots}
\pgfplotsset{compat=newest}
\usepgfplotslibrary{groupplots}
\usepgfplotslibrary{dateplot}
\usepackage{multirow} 
\usepackage{booktabs} 
\usepackage{array}
\usetikzlibrary{arrows,arrows.meta,automata,shapes,calc,intersections}
\usepackage{capt-of}
\usepackage[ruled,vlined,linesnumbered]{algorithm2e}

\definecolor{color1}{HTML}{6FADA7}
\definecolor{color2}{HTML}{B04550}
\newcommand{\highlight}[1]{\textcolor{color1}{#1}}
\newcommand{\highlightb}[1]{\textcolor{color2}{#1}}

\IEEEoverridecommandlockouts
\begin{document}
\newcommand{\matA}{\mathbf{A}}
\newcommand{\bfA}{\mathbf{A}}
\newcommand{\calA}{\mathcal{A}}
\newcommand{\bbA}{\mathbb{A}}
\newcommand{\bfa}{\mathbf{a}}
\newcommand{\veca}{\mathbf{a}}
\newcommand{\matB}{\mathbf{B}}
\newcommand{\bfB}{\mathbf{B}}
\newcommand{\calB}{\mathcal{B}}
\newcommand{\bbB}{\mathbb{B}}
\newcommand{\bfb}{\mathbf{b}}
\newcommand{\vecb}{\mathbf{b}}
\newcommand{\matC}{\mathbf{C}}
\newcommand{\bfC}{\mathbf{C}}
\newcommand{\calC}{\mathcal{C}}
\newcommand{\bbC}{\mathbb{C}}
\newcommand{\bfc}{\mathbf{c}}
\newcommand{\vecc}{\mathbf{c}}
\newcommand{\matD}{\mathbf{D}}
\newcommand{\bfD}{\mathbf{D}}
\newcommand{\calD}{\mathcal{D}}
\newcommand{\bbD}{\mathbb{D}}
\newcommand{\bfd}{\mathbf{d}}
\newcommand{\vecd}{\mathbf{d}}
\newcommand{\matE}{\mathbf{E}}
\newcommand{\bfE}{\mathbf{E}}
\newcommand{\calE}{\mathcal{E}}
\newcommand{\bbE}{\mathbb{E}}
\newcommand{\bfe}{\mathbf{e}}
\newcommand{\vece}{\mathbf{e}}
\newcommand{\matF}{\mathbf{F}}
\newcommand{\bfF}{\mathbf{F}}
\newcommand{\calF}{\mathcal{F}}
\newcommand{\bbF}{\mathbb{F}}
\newcommand{\bff}{\mathbf{f}}
\newcommand{\vecf}{\mathbf{f}}
\newcommand{\matG}{\mathbf{G}}
\newcommand{\bfG}{\mathbf{G}}
\newcommand{\calG}{\mathcal{G}}
\newcommand{\bbG}{\mathbb{G}}
\newcommand{\bfg}{\mathbf{g}}
\newcommand{\vecg}{\mathbf{g}}
\newcommand{\matH}{\mathbf{H}}
\newcommand{\bfH}{\mathbf{H}}
\newcommand{\calH}{\mathcal{H}}
\newcommand{\bbH}{\mathbb{H}}
\newcommand{\bfh}{\mathbf{h}}
\newcommand{\vech}{\mathbf{h}}
\newcommand{\matI}{\mathbf{I}}
\newcommand{\bfI}{\mathbf{I}}
\newcommand{\calI}{\mathcal{I}}
\newcommand{\bbI}{\mathbb{I}}
\newcommand{\bfi}{\mathbf{i}}
\newcommand{\veci}{\mathbf{i}}
\newcommand{\matJ}{\mathbf{J}}
\newcommand{\bfJ}{\mathbf{J}}
\newcommand{\calJ}{\mathcal{J}}
\newcommand{\bbJ}{\mathbb{J}}
\newcommand{\bfj}{\mathbf{j}}
\newcommand{\vecj}{\mathbf{j}}
\newcommand{\matK}{\mathbf{K}}
\newcommand{\bfK}{\mathbf{K}}
\newcommand{\calK}{\mathcal{K}}
\newcommand{\bbK}{\mathbb{K}}
\newcommand{\bfk}{\mathbf{k}}
\newcommand{\veck}{\mathbf{k}}
\newcommand{\matL}{\mathbf{L}}
\newcommand{\bfL}{\mathbf{L}}
\newcommand{\calL}{\mathcal{L}}
\newcommand{\bbL}{\mathbb{L}}
\newcommand{\bfl}{\mathbf{l}}
\newcommand{\vecl}{\mathbf{l}}
\newcommand{\matM}{\mathbf{M}}
\newcommand{\bfM}{\mathbf{M}}
\newcommand{\calM}{\mathcal{M}}
\newcommand{\bbM}{\mathbb{M}}
\newcommand{\bfm}{\mathbf{m}}
\newcommand{\vecm}{\mathbf{m}}
\newcommand{\matN}{\mathbf{N}}
\newcommand{\bfN}{\mathbf{N}}
\newcommand{\calN}{\mathcal{N}}
\newcommand{\bbN}{\mathbb{N}}
\newcommand{\bfn}{\mathbf{n}}
\newcommand{\vecn}{\mathbf{n}}
\newcommand{\matO}{\mathbf{O}}
\newcommand{\bfO}{\mathbf{O}}
\newcommand{\calO}{\mathcal{O}}
\newcommand{\bbO}{\mathbb{O}}
\newcommand{\bfo}{\mathbf{o}}
\newcommand{\veco}{\mathbf{o}}
\newcommand{\matP}{\mathbf{P}}
\newcommand{\bfP}{\mathbf{P}}
\newcommand{\calP}{\mathcal{P}}
\newcommand{\bbP}{\mathbb{P}}
\newcommand{\bfp}{\mathbf{p}}
\newcommand{\vecp}{\mathbf{p}}
\newcommand{\matQ}{\mathbf{Q}}
\newcommand{\bfQ}{\mathbf{Q}}
\newcommand{\calQ}{\mathcal{Q}}
\newcommand{\bbQ}{\mathbb{Q}}
\newcommand{\bfq}{\mathbf{q}}
\newcommand{\vecq}{\mathbf{q}}
\newcommand{\matR}{\mathbf{R}}
\newcommand{\bfR}{\mathbf{R}}
\newcommand{\calR}{\mathcal{R}}
\newcommand{\bbR}{\mathbb{R}}
\newcommand{\bfr}{\mathbf{r}}
\newcommand{\vecr}{\mathbf{r}}
\newcommand{\matS}{\mathbf{S}}
\newcommand{\bfS}{\mathbf{S}}
\newcommand{\calS}{\mathcal{S}}
\newcommand{\bbS}{\mathbb{S}}
\newcommand{\bfs}{\mathbf{s}}
\newcommand{\vecs}{\mathbf{s}}
\newcommand{\matT}{\mathbf{T}}
\newcommand{\bfT}{\mathbf{T}}
\newcommand{\calT}{\mathcal{T}}
\newcommand{\bbT}{\mathbb{T}}
\newcommand{\bft}{\mathbf{t}}
\newcommand{\vect}{\mathbf{t}}
\newcommand{\matU}{\mathbf{U}}
\newcommand{\bfU}{\mathbf{U}}
\newcommand{\calU}{\mathcal{U}}
\newcommand{\bbU}{\mathbb{U}}
\newcommand{\bfu}{\mathbf{u}}
\newcommand{\vecu}{\mathbf{u}}
\newcommand{\matV}{\mathbf{V}}
\newcommand{\bfV}{\mathbf{V}}
\newcommand{\calV}{\mathcal{V}}
\newcommand{\bbV}{\mathbb{V}}
\newcommand{\bfv}{\mathbf{v}}
\newcommand{\vecv}{\mathbf{v}}
\newcommand{\matW}{\mathbf{W}}
\newcommand{\bfW}{\mathbf{W}}
\newcommand{\calW}{\mathcal{W}}
\newcommand{\bbW}{\mathbb{W}}
\newcommand{\bfw}{\mathbf{w}}
\newcommand{\vecw}{\mathbf{w}}
\newcommand{\matX}{\mathbf{X}}
\newcommand{\bfX}{\mathbf{X}}
\newcommand{\calX}{\mathcal{X}}
\newcommand{\bbX}{\mathbb{X}}
\newcommand{\bfx}{\mathbf{x}}
\newcommand{\vecx}{\mathbf{x}}
\newcommand{\matY}{\mathbf{Y}}
\newcommand{\bfY}{\mathbf{Y}}
\newcommand{\calY}{\mathcal{Y}}
\newcommand{\bbY}{\mathbb{Y}}
\newcommand{\bfy}{\mathbf{y}}
\newcommand{\vecy}{\mathbf{y}}
\newcommand{\matZ}{\mathbf{Z}}
\newcommand{\bfZ}{\mathbf{Z}}
\newcommand{\calZ}{\mathcal{Z}}
\newcommand{\bbZ}{\mathbb{Z}}
\newcommand{\bfz}{\mathbf{z}}
\newcommand{\vecz}{\mathbf{z}}
\newcommand{\bfalpha}{\boldsymbol{\alpha}}
\newcommand{\vecalpha}{\boldsymbol{\alpha}}
\newcommand{\bfbeta}{\boldsymbol{\beta}}
\newcommand{\vecbeta}{\boldsymbol{\beta}}
\newcommand{\bfgamma}{\boldsymbol{\gamma}}
\newcommand{\vecgamma}{\boldsymbol{\gamma}}
\newcommand{\bfdelta}{\boldsymbol{\delta}}
\newcommand{\vecdelta}{\boldsymbol{\delta}}
\newcommand{\bfepsilon}{\boldsymbol{\epsilon}}
\newcommand{\vecepsilon}{\boldsymbol{\epsilon}}
\newcommand{\bfzeta}{\boldsymbol{\zeta}}
\newcommand{\veczeta}{\boldsymbol{\zeta}}
\newcommand{\bfeta}{\boldsymbol{\eta}}
\newcommand{\veceta}{\boldsymbol{\eta}}
\newcommand{\bftheta}{\boldsymbol{\theta}}
\newcommand{\vectheta}{\boldsymbol{\theta}}
\newcommand{\bfiota}{\boldsymbol{\iota}}
\newcommand{\veciota}{\boldsymbol{\iota}}
\newcommand{\bfkappa}{\boldsymbol{\kappa}}
\newcommand{\veckappa}{\boldsymbol{\kappa}}
\newcommand{\bflamda}{\boldsymbol{\lamda}}
\newcommand{\veclamda}{\boldsymbol{\lamda}}
\newcommand{\bfmu}{\boldsymbol{\mu}}
\newcommand{\vecmu}{\boldsymbol{\mu}}
\newcommand{\bfnu}{\boldsymbol{\nu}}
\newcommand{\vecnu}{\boldsymbol{\nu}}
\newcommand{\bfxi}{\boldsymbol{\xi}}
\newcommand{\vecxi}{\boldsymbol{\xi}}
\newcommand{\bfomicron}{\boldsymbol{\omicron}}
\newcommand{\vecomicron}{\boldsymbol{\omicron}}
\newcommand{\bfpi}{\boldsymbol{\pi}}
\newcommand{\vecpi}{\boldsymbol{\pi}}
\newcommand{\bfrho}{\boldsymbol{\rho}}
\newcommand{\vecrho}{\boldsymbol{\rho}}
\newcommand{\bfsigma}{\boldsymbol{\sigma}}
\newcommand{\vecsigma}{\boldsymbol{\sigma}}
\newcommand{\bftau}{\boldsymbol{\tau}}
\newcommand{\vectau}{\boldsymbol{\tau}}
\newcommand{\bfupsilon}{\boldsymbol{\upsilon}}
\newcommand{\vecupsilon}{\boldsymbol{\upsilon}}
\newcommand{\bfphi}{\boldsymbol{\phi}}
\newcommand{\vecphi}{\boldsymbol{\phi}}
\newcommand{\bfchi}{\boldsymbol{\chi}}
\newcommand{\vecchi}{\boldsymbol{\chi}}
\newcommand{\bfpsi}{\boldsymbol{\psi}}
\newcommand{\vecpsi}{\boldsymbol{\psi}}
\newcommand{\bfomega}{\boldsymbol{\omega}}
\newcommand{\vecomega}{\boldsymbol{\omega}}
\newcommand{\bfAlpha}{\boldsymbol{\Alpha}}
\newcommand{\matAlpha}{\boldsymbol{\Alpha}}
\newcommand{\bfBeta}{\boldsymbol{\Beta}}
\newcommand{\matBeta}{\boldsymbol{\Beta}}
\newcommand{\bfGamma}{\boldsymbol{\Gamma}}
\newcommand{\matGamma}{\boldsymbol{\Gamma}}
\newcommand{\bfDelta}{\boldsymbol{\Delta}}
\newcommand{\matDelta}{\boldsymbol{\Delta}}
\newcommand{\bfEpsilon}{\boldsymbol{\Epsilon}}
\newcommand{\matEpsilon}{\boldsymbol{\Epsilon}}
\newcommand{\bfZeta}{\boldsymbol{\Zeta}}
\newcommand{\matZeta}{\boldsymbol{\Zeta}}
\newcommand{\bfEta}{\boldsymbol{\Eta}}
\newcommand{\matEta}{\boldsymbol{\Eta}}
\newcommand{\bfTheta}{\boldsymbol{\Theta}}
\newcommand{\matTheta}{\boldsymbol{\Theta}}
\newcommand{\bfIota}{\boldsymbol{\Iota}}
\newcommand{\matIota}{\boldsymbol{\Iota}}
\newcommand{\bfKappa}{\boldsymbol{\Kappa}}
\newcommand{\matKappa}{\boldsymbol{\Kappa}}
\newcommand{\bfLamda}{\boldsymbol{\Lamda}}
\newcommand{\matLamda}{\boldsymbol{\Lamda}}
\newcommand{\bfMu}{\boldsymbol{\Mu}}
\newcommand{\matMu}{\boldsymbol{\Mu}}
\newcommand{\bfNu}{\boldsymbol{\Nu}}
\newcommand{\matNu}{\boldsymbol{\Nu}}
\newcommand{\bfXi}{\boldsymbol{\Xi}}
\newcommand{\matXi}{\boldsymbol{\Xi}}
\newcommand{\bfOmicron}{\boldsymbol{\Omicron}}
\newcommand{\matOmicron}{\boldsymbol{\Omicron}}
\newcommand{\bfPi}{\boldsymbol{\Pi}}
\newcommand{\matPi}{\boldsymbol{\Pi}}
\newcommand{\bfRho}{\boldsymbol{\Rho}}
\newcommand{\matRho}{\boldsymbol{\Rho}}
\newcommand{\bfSigma}{\boldsymbol{\Sigma}}
\newcommand{\matSigma}{\boldsymbol{\Sigma}}
\newcommand{\bfTau}{\boldsymbol{\Tau}}
\newcommand{\matTau}{\boldsymbol{\Tau}}
\newcommand{\bfUpsilon}{\boldsymbol{\Upsilon}}
\newcommand{\matUpsilon}{\boldsymbol{\Upsilon}}
\newcommand{\bfPhi}{\boldsymbol{\Phi}}
\newcommand{\matPhi}{\boldsymbol{\Phi}}
\newcommand{\bfChi}{\boldsymbol{\Chi}}
\newcommand{\matChi}{\boldsymbol{\Chi}}
\newcommand{\bfPsi}{\boldsymbol{\Psi}}
\newcommand{\matPsi}{\boldsymbol{\Psi}}
\newcommand{\bfOmega}{\boldsymbol{\Omega}}
\newcommand{\matOmega}{\boldsymbol{\Omega}}
\newcommand{\bfzero}{\boldsymbol{0}}

\newcommand{\card}[1]{{|{#1}|}}
\newcommand{\abs}[1]{{\left|{#1}\right|}}
\newcommand{\sgn}[1]{{\text{sgn}\left({#1}\right)}}
\newcommand{\norm}[2][]{{\left\Vert{#2}\right\Vert}_{#1}}
\newcommand{\inprod}[3][]{{\left\langle{#2},{#3}\right\rangle}_{#1}}
\newcommand{\dotp}[3][]{{\left\langle{#2},{#3}\right\rangle}_{#1}}
\newcommand{\set}[1]{{\{#1\}}}
\newcommand{\trace}[1]{\text{trace}\left(#1\right)}
\newcommand{\argmin}{\mathop{\text{argmin}}}
\newcommand{\argmax}{\mathop{\text{argmax}}}
\newcommand{\intseq}[2]{[{#1};{#2}]}
\newcommand{\eqdef}{\triangleq}
\newcommand{\E}[2][]{{\mathbb{E}_{#1}\left[#2\right]}}
\newcommand{\Var}[1]{{\text{Var}\left(#1\right)}}
\newcommand{\Cov}[1]{{\text{Cov}\left(#1\right)}}
\renewcommand{\P}[2][]{{\mathbb{P}_{#1}\left(#2\right)}}
\newcommand{\indic}[1]{{\mathbf{1}\{#1\}}}
\newcommand{\mat}[2]{\left[\begin{array}{#1}#2\end{array}\right]}
\newcommand{\bra}[1]{{\left\langle{#1}\right\vert}}
\newcommand{\ket}[1]{{\left\vert{#1}\right\rangle}}

\newcommand{\D}[3][]{{\mathbb{D}_{#1}}\!\left(#2\,\middle\Vert\,#3\right)} % divergence
\newcommand{\V}[3][]{{\mathbb{V}_{#1}}\!\left(#2,#3\right)} % total variation
\newcommand{\I}[2]{{{\mathbb{I}}\!\left(#1;#2\right)}} % mutual information (two params)
\newcommand{\avgI}[1]{{{\mathbb{I}}\!\left(#1\right)}} % mutual information
\newcommand{\avgH}[1]{{\mathbb{H}}\!\left(#1\right)}
\newcommand{\h}[1]{{{h}}\left(#1\right)} % differential entropy
\newcommand{\Hb}[1]{{h_b}\left(#1\right)} % Binary entropy function

\newcommand{\GF}[1]{\ensuremath{\mbox{GF}(#1)}} % Galois field
\newcommand{\F}[1]{\ensuremath{\mathbb{F}_{#1}}} % Field
\newcommand{\dH}[1]{\ensuremath{\mbox{d}_{\mbox{\tiny H}}(#1)}} % Hamming distance
\newcommand{\wt}[1]{\ensuremath{\mbox{wt}(#1)}} % Weight of a vector

\newcommand{\dd}{\text{\textnormal{d}}} % differential operator as in dx
\renewcommand{\leq}{\leqslant} % redefines <=
\renewcommand{\geq}{\geqslant} % redefines >=

% Matrix operations
\newcommand{\tr}[2][]{\ensuremath{\text{\textnormal{tr}}_{#1}\left(#2\right)}}  % Trace
\newcommand{\rk}[1]{\ensuremath{\text{\textnormal{rank}}\left(#1\right)}}  % Rank
\renewcommand{\det}[1]{{\left|{#1}\right|}}                              % Det
\renewcommand{\ker}[1]{\ensuremath{\text{\textnormal{Ker}}\left(#1\right)}}  % Kernel

\newcommand{\cn}{\textcolor{red}{[\raisebox{-0.2ex}{\tiny\shortstack{citation\\[-1ex]needed}}]}\xspace}
\acrodef{ACDIS}[ACDIS]{Adaptive Communication Decision and Information Systems}
\acrodef{AWGN}{additive white Gaussian noise}
\acrodef{BER}{bit error rate}
\acrodef{BEC}{binary erasure channel}
\acrodef{BPSK}{binary phase-shift keying}
\acrodef{BSC}{binary symmetric channel}
\acrodef{BM}[BM]{branch metric}
\acrodef{CDF}[CDF]{cumulative distribution function}
\acrodef{CDEGA}[CDEGA]{conditional DEGA}
\acrodef{CLT}[CLT]{central limit theorem}
\acrodef{DMC}[DMC]{discrete memoryless channel}
\acrodef{DMS}[DMS]{discrete memoryless source}
\acrodef{DEGA}[DEGA]{density evolution Gaussian approximation}
\acrodef{eBCH}[eBCH]{extended Bose–Chaudhuri–Hocquenghem}
\acrodef{FER}[FER]{frame error rate}
\acrodef{iid}[i.i.d.]{independent and identically distributed}
\acrodef{IoT}[IoT]{Internet of Things}
\acrodef{ISIT}[ISIT]{the International Symposium of Information Theory}
\acrodef{LDPC}[LDPC]{low-density parity-check}
\acrodef{LLR}[LLR]{log-likelihood ratio}
\acrodef{MAC}[MAC]{multiple-access channel}
\acrodef{ML}[ML]{maximum likelihood}
\acrodef{MLC}[MLC]{multi-level coding}
\acrodef{MLE}[MLE]{maximum likelihood estimate}
\acrodef{MIMO}[MIMO]{multiple-input multiple-output}
\acrodef{MSD}[MSD]{multi-stage decoding}
\acrodef{PAC}[PAC]{polarization-adjusted convolutional}
\acrodef{PDF}[PDF]{probability density function}
\acrodef{PM}[PM]{path metric}
\acrodef{PMF}[PMF]{probability mass function}
\acrodef{PPM}[PPM]{pulse position modulation}
\acrodef{PSK}{phase shift keying}
\acrodef{URLLC}[URLLC]{Ultra-Reliable Low-Latency Communication}
\acrodef{QKD}{quantum key distribution}
\acrodef{QPSK}{quadrature phase-shift keying}
\acrodef{RM}[RM]{Reed-Muller}
\acrodef{RV}{random variable}
\acrodef{SCD}[SCD]{successive cancellation decoding}
\acrodef{SCL}[SCL]{successive cancellation list}
\acrodef{SC}[SC]{successive cancellation}
\acrodef{SNR}{signal-to-noise ratio}
\acrodef{wrt}[w.r.t.]{with respect to}
\acrodef{OSD}[OSD]{ordered-statistics decoding}
\acrodef{GRAND}[GRAND]{guessing random additive noise decoding}
%\title{A Path Metric Based Construction of Polarization-Adjusted Convolutional Codes}
\title{An SCL-Aware Design of PAC Codes}
%%%%%%
\author{%
  \IEEEauthorblockN{Tyler Kann$^1$, Shrinivas Kudekar, Matthieu R. Bloch$^1$}  \thanks{
This work was supported in part by the National Science Foundation (NSF) under grant 2148400 as part of the Resilient $\&$ Intelligent NextG Systems (RINGS) program. The work is a continuation of work presented at ISIT in 2024 \cite{kannPathMetricBased2024}.}
  \IEEEauthorblockA{%
  $^1$ School of Electrical and Computer Engineering, Georgia Institute of Technology, Atlanta, GA 30332, USA \\
  kann@gatech.edu, kudekar@gmail.com, matthieu.bloch@ece.gatech.edu }
}

\maketitle

\begin{abstract}
   We develop design methodologies for the rate profile of \ac{PAC} codes tailored to \ac{SCL} decoding. The crux of our approach is two-fold: first is the creation of a modified \ac{DEGA}, called \ac{CDEGA}, to analyze the decoding process in the presence of errors; second is the creation of a dangerous-path occupancy metric to find the conditional occupancy of bit-errors in a finite list. Our approach not only illuminates the role of rate-one precoding, but also  provides insight into why certain information sets are particularly well suited to \ac{PAC} codes. We demonstrate the usefulness of \ac{CDEGA} and dangerous-path metrics with algorithms that create new information sets, as well as precoders that are extremely sparse yet competitive. In particular, we design codes of length $N \leq 512$ with various rates and list sizes, all of which offer at least half a dB gain over traditional \ac{PAC} codes at a \ac{FER} of $10^{-3}$ and improve upon state-of-the-art results.
\end{abstract}
  %\ac{PAC} codes, proposed in 2019 by Arikan at \ac{ISIT}, are a rate 1 precoding that is applied before the polar transform. This was shown to enhance polarization at shorter blocklengths and substantially improve block error rate for channel codes. We investigate the application of \ac{PAC} to source coding with side information, and show how the algorithm can be modified to accommodate the view of source coding as the reverse of channel coding. We show empirical results as comparisons with traditional list decoding of polar codes. Benefits to secret key generation is briefly discussed.        

\section{Introduction}
\label{sec:intro}

A key objective of 6G is to achieve communication with improved reliability and less latency than previous standards, formalized with the \ac{URLLC} requirement. \ac{URLLC} demands new and improved error-control coding for extremely short blocklengths, typically $N \leq 256$ \cite{Yue2022}. Promising coding schemes that have shown strong performance include \ac{eBCH}~\cite{eBCH}; \ac{RM} codes~\cite{Coskun2020Successive}; polar codes~\cite{Arikan2009} and \acf{PAC} codes~\cite{ArikanPAC,RowshanViterbi,Rowshan2021Polarization}, especially when used in combination with decoders with memory such as \acf{SCL} decoding ~\cite{Rowshan2021Polarization, VardyListPAC}; \acf{OSD}~\cite{OSD,yueOrderedStatisticsDecodingAdaptive2022, Cheng_Chen_2021}; or \acf{GRAND}~\cite{Duffy2019Capacity, GRAND_SO}.

While earlier works have exploited the benefits of PAC codes for large improvements at short blocklengths, there have been few attempts to develop a principled explanation for the boost in performance PAC provides over traditional polar coding. Existing attempts have focused on cutoff metrics \cite{WeightedSum}, a joint source-coding perspective \cite{sunAnalysisPolarizationadjustedConvolutional2021}, and primarily on the impact the precoder has on the weight spectrum of the code \cite{rowshanFormationMinweightCodewords2023,jiangConstructionPACCodes2023}. However, none of the current methods in the literature directly tie the performance of \ac{PAC} to the decoder, or more explicitly, why PAC benefits are not realizable under \ac{SCD} but are attained under \ac{SCL} or Fano decoders that track multiple paths. While \ac{SCL} can approximate \acf{ML} performance, a regime where the weight spectrum dominates performance, the list size at which this approximation is accurate may be too large for a practical design. Our key observation is that finite-list rate profiling requires an additional statistics describing when and how decoding errors occur in the decoding process. 

The main contribution of the present work is to design rate profiles in a way that is explicitly tied to \ac{SCL}. We achieve this by first analyzing how \ac{PAC} codes influence the \ac{PM}, which offers insight into why a precoder benefits \ac{SCL}. We then introduce a tool to determine which bit channels are most influential in the \ac{FER}, which we use in our algorithm to create improved information sets.

The remainder of the paper is organized as follows. In Section \ref{sec:pac-codes-intro}, we review necessary notation and concepts regarding polar codes, \ac{DEGA}, and \ac{PAC}. In Section \ref{sec:cdega}, we introduce \acf{CDEGA}, a tool to analyze the behavior of PAC codes, as well as an algorithm to construct \ac{CDEGA} based information sets. In Section \ref{sec:tail-events}, we introduce data driven metrics, primarily what we call the dangerous-path spectrum, to improve upon \ac{CDEGA}, as well as algorithms to construct information sets and precoders. In Section~\ref{sec:results}, we present numerical results and conclude in Section \ref{sec:conc}. 
\section{Background and notation}
\label{sec:pac-codes-intro}

\subsection{Polar codes} 
\label{sec:polar-codes}

A polar code for channel coding is characterized by its blocklength $N \eqdef 2^n, n \in \mathbb{N}$, the number of information bits $K$, and an information set $\mathcal{A}\subset\intseq{0}{N-1}$ that specifies where to place information bits. Specifically, a vector of $K$ information bits $m$  is encoded into a length $N$ vector $u$ such that $u_\mathcal{A} \eqdef m$ and $u_{\calF} \eqdef 0$, the all-zero vector. The sets $\calA$ and $\calF$ are called the information and frozen set, respectively. Upon setting $G_2 = \left[\begin{smallmatrix} 1 & 0 \\ 1 & 1 \end{smallmatrix}\right]$,  the base polarization matrix, and $G^{\otimes n}$ as the $n$th order Kronecker product of $G_2$, a codeword $x$ is created through the operation $x \eqdef uG^{\otimes n}$, a process known as the polar transform. The structure of the matrix $G^{\otimes n}$ allows for an encoding complexity of $\mathcal{O}(N\log N)$. The codeword $x$ is then sent over the channel to a receiver that observes a noisy version $y$, e.g., $y_i= x_i+ n_i, n_i \sim \calN(0,\sigma^2)$ if each symbol $x_i$ of the codeword $x$ is transmitted through an \acf{AWGN} channel. 
  
The standard decoding algorithm for polar codes is the \ac{SC} decoder, by which bits from the information set $\mathcal{A}$ are successively decoded based on their \ac{LLR}, $\lambda$, given the past decoded bits according to the maximum-likelihood rule. 
\begin{align}
\forall i\in\calA\phantom{^c}\quad  \hat{u}^0_i &=     \begin{cases}
                    0 & \text{if } \lambda^0_i = \ln \frac{P(y,\hat{u}_{0:i-1}|\hat{u}^0_i = 0) }{P(y,\hat{u}_{0:i-1}|\hat{u}^0_i = 1) } > 0 \\
                    1 & \text{else.} \label{info}
                                         \end{cases},\\
  \forall i\in\calA^c\quad \hat{u}^0_i&=0,
\end{align}
where $\hat{u}_{0:i-1} = \{\hat{u}^0_0, \hat{u}^0_1, \dots, \hat{u}^0_{i-1} \}$ denotes the vector of past decisions. The \acp{LLR} can be efficiently computed recursively, resulting in a decoding complexity of $\mathcal{O}(N\log N)$. The \acp{LLR} can also be viewed as decisions made at the output of individual \emph{bit channels}, corresponding to channels with input bit $u_i$ and output $(y,u_{0:i-1})$. The choice of the set $\calA$, called \emph{rate profiling}, plays a crucial role in determining the performance of polar codes. Several criteria have been proposed based on the capacity of the bit channels~\cite{Tal_2013} or the \ac{RM} profile~\cite{RMPolar}, which consists in enumerating the Hamming weight of \ac{RM} codewords and selecting those with highest weight.

\subsection{Successive cancellation list decoding}
\label{sec:SCL}

Since each decision in the \ac{SC} decoder relies on previously decoded bits, a single bad decision can propagate and make future bits erroneous. \ac{SCL} decoding~\cite{Tal_2015} offers an efficient solution to avoid this problem. In \ac{SCL}, the decoding process is viewed as following branches of a tree, simultaneously tracking up to $L$ branches. The tree splits at every information bit $i$, creating two branches: $\hat{u}^0_i = 0$ and $\hat{u}^0_i = 1$. For every path in the set $\calL$ of currently tracked paths, the \ac{PM} is computed as the accumulation of \acp{BM}. Each \ac{BM} is a penalty based on the decision for both frozen and information bits, according to: 
\begin{align}
&\forall l\in\calL, \quad  \textrm{PM}_{i,\ell} = \textrm{PM}_{i-1,\ell} + \textrm{BM}_{i,\ell} \\ 
& \textrm{BM}_{i,\ell} =
\begin{cases}
                    0 & \text{if } 1 - 2\hat{u}^0_{i,\ell} = \textrm{sign}(\lambda^0_{i,\ell})  \\
                    |\lambda^0_{i,\ell}| & \text{if } 1 - 2\hat{u}^0_{i,\ell} = -\textrm{sign}(\lambda^0_{i,\ell}) 
    \end{cases},
    \label{BM}
\end{align} 
where the subscript $(i,\ell)$ denotes bit $i$ on path $\ell$. When $|\calL|$ is greater than $L$, the list is pruned back to only $L$ paths, keeping those with the lowest path metrics. 

In list decoding, there are two possible error events that result in a decoding failure. The first error event occurs when the paths are pruned from $2L$ to $L$, and the correct path is deleted in this process. The second error occurs at the end, when the correct path is one of the $L$ remaining, but not the path with the lowest \ac{PM}. We refer to these errors as path-loss errors and selection errors, respectively. The selection error occurs when there exists at least one path with a smaller PM than the correct path. We refer to the incorrect paths with path metrics smaller than the correct path as \textit{dangerous paths}, and to the smallest path metric of these paths as the most misleading path metric (MMPM). The dangerous paths are the paths that determines if there is a selection error, which happens if and only if the MMPM is less than the correct \ac{PM}. Thus one possible guiding principle for code design is to increase the value of the MMPM and consequently reduce the number of selection errors.

\subsection{PAC codes}
\label{sec:pac-codes}

While polar codes are asymptotically optimal \cite{Arikan2009}, they are not immediately competitive in the short to medium length regime. The polarization process, by which channels eventually become either completely noisy or noiseless, happens relatively slowly, and smaller length codes contain channels that do not yet fully polarize. To improve the performance in the short to medium length regime, several modifications of polar codes have been proposed. Successful designs include cyclic redundancy checks and \acf{PAC} codes, both aiming to aid the decoder. \ac{PAC} codes, introduced in \cite{ArikanPAC}, can be viewed as polar codes with dynamically frozen bits \cite{Trifonov2013,Trifonov2020Randomized}. Surprisingly, precoding the polar code with an outer \emph{rate-one} convolutional code offers significant improvements in performance with no rate penalty. 

A precoded polar code still polarizes $u$ to get the final codeword, except now $u = vT$, with $v$ as the new information vector and $T$ representing a causal, upper-triangular precoding matrix. The information set $\mathcal{A}$ now refers to $v$, with $v_{\mathcal{A}} = m, v_{\mathcal{F}} = 0$. Decoding and metric decisions from \mbox{~\eqref{info}-\eqref{BM}} are still based on $\hat{u}_i$, but the decoder's branches are based on $\hat{v}_i$ and $\hat{u}_i$ is now determined from $\hat{v}_{0:i}$ and $T$.

\subsection{Density evolution Gaussian approximation}
\label{sec:dega}

The key challenge when designing polar codes and their modifications is creating the information set $\mathcal{A}$ that minimizes the block error probability. This reduces to choosing information bits that are deemed the most reliable according to the chosen metric of channel reliability. One method for estimating the reliabilities of polarized channels is \acf{DEGA}, based on \cite{chung_analysis_2001} and applied to polar codes in \cite{moriPerformancePolarCodes2009,Trifonov2013}. The received \acp{LLR} are treated as Gaussian distributions corresponding to the all-zero codeword, parameterized as $\mathcal{N}(\frac{2}{\sigma^2_N},\frac{4}{\sigma^2_N})$, where $\sigma^2_N$ is the noise variance of the channel. We aim to calculate the \emph{posterior} distributions of the \acp{LLR} at the output of the individual bit channels.  The specific realizations of the \acp{LLR} are denoted as $\lambda$, and the means of the LLRs are denoted as $\mu$. Since the variance is set to twice the mean, a property referred to as symmetry, the entire distribution can be characterized solely by $\mu$. The distributions then evolve through the polarization tree, giving the following recursive calculations of $\mu$:
\begin{align}
\mu_{i}^{k} &= \phi^{-1}\Big(1-\big(1-\phi(\mu_{i}^{k+1})\big) \nonumber \\ &\phantom{......} \times \big(1-\phi(\mu_{i+2^{k}}^{k+1})\big)\Big), \qquad \textrm{bin}(i)_k=0 \label{eq:checknode}\\
\mu_{i}^{k} &= \mu_{i}^{k+1} + \mu_{i-2^{k}}^{k+1}, \phantom{.}\qquad\qquad\ \, \textrm{bin}(i)_k=1 \label{eq:repitition},
\end{align}
where $\phi$ is the DEGA approximation function from \cite{chung_analysis_2001}, and $\textrm{bin}(i) \triangleq B_{n-1}B_{n-2}\dots B_{1}B_0$ is the $n$-bit binary representation of  $i$. 

The evolution of the mean through check nodes and variable nodes of the polar process is captured by \eqref{eq:checknode} and \eqref{eq:repitition}. In an actual decoder run, repetition nodes may actually subtract \acp{LLR} if the previous bit decision was a $1$. However, since \ac{DEGA} tracks the all-zero codeword, \emph{all} repetition nodes correspond to addition of LLRs. Compared to low-density parity-check codes, polar codes form a perfect tree with no cycles and exactly two inputs in the density evolution. This makes the posteriors not only easier to calculate but also accurate, which can be used to predict the decoding performance.

\section{Conditional DEGA}
\label{sec:cdega}

 DEGA considers the transmission of the all-zero codeword, and the reliability of each bit is based on the assumption that all prior bits have been decoded correctly. This assumption is sufficient to analyze \ac{SCD}, in which one cannot afford to make any error.  However, the improved performance of decoders that track multiple paths (Fano, Stack, SCL, etc.) motivates us to investigate what happens to posteriors conditioned on an error made earlier. 

\begin{figure}[htb]
  \begin{subfigure}{.5\textwidth}
    \centering
   \includegraphics[width=\textwidth]{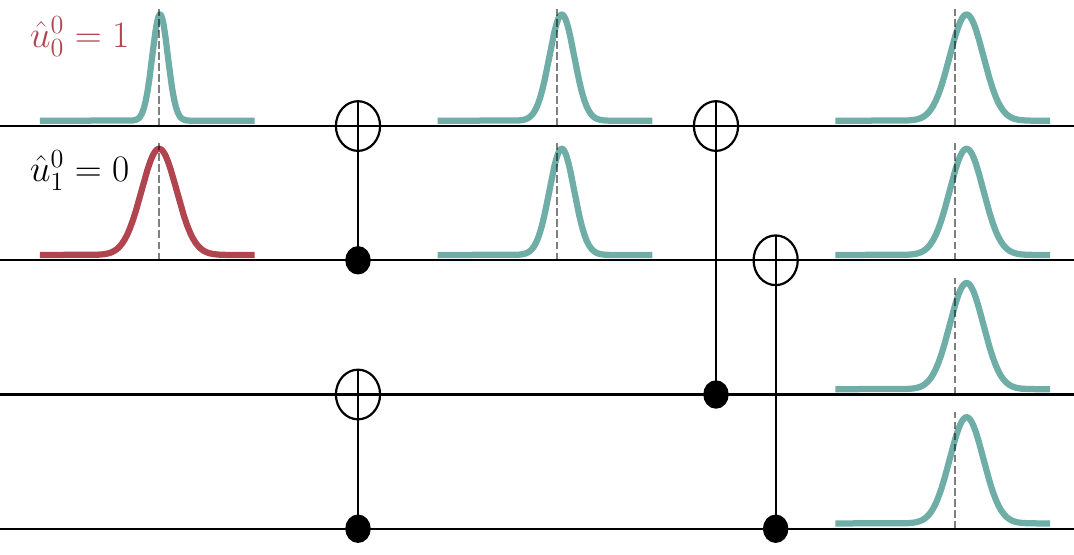}
   %\vspace{-8mm}
    \caption{The received \acp{LLR} are on the ``right'' and the posterior \acp{LLR} are computed on the ``left''. In traditional \ac{DEGA}, the distributions are symmetric and  shrink towards zero when going through a check node and move towards $+\infty$ when going through a repetition node. Because of the distribution symmetry, as the distributions move towards zero, the variance also shrinks to zero. We assume here that bit $u_0^0$ is erroneously decoded to $1$. This causes subtraction on the repetition node, and the resulting distribution of the posterior for bit $u_1^0$ has zero mean. The correct distributions are turquoise, whereas distributions affected by~\eqref{eq:subtraction} are in red.}
  \end{subfigure}
  \hspace{1px}
  \begin{subfigure}{.5\textwidth}
    \centering
   \vspace{2mm}
   \includegraphics[width=\textwidth]{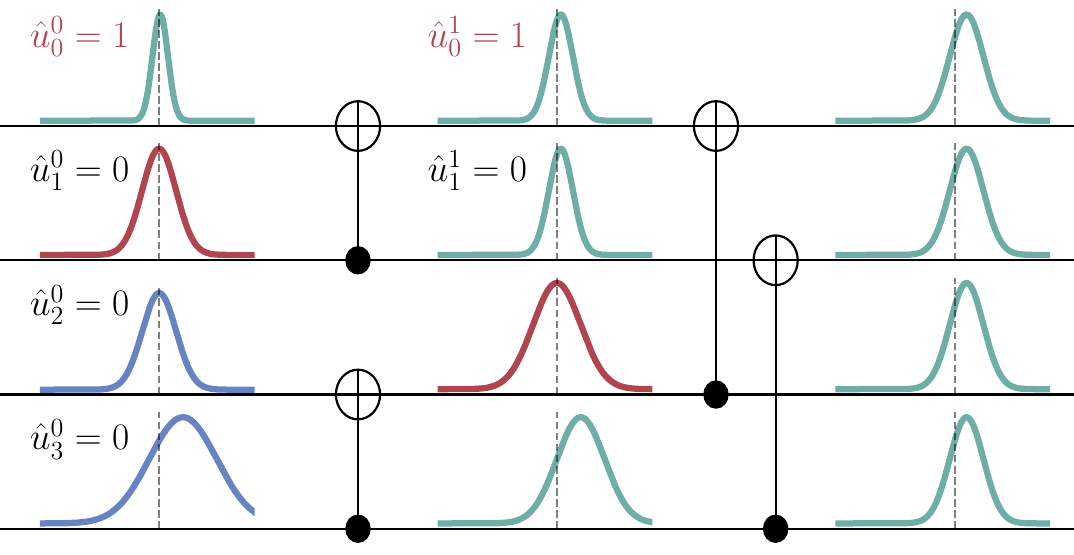}
    \caption{Because of the erroneous $1$, the second layer polarizes to $(1,0)$. This causes subtraction amongst the received \acp{LLR}. The distribution of the posterior on $u_2^0$ is zero mean but not zero variance; the posterior of $u_3^0$ is distributed by $\calN(2\mu,8\mu)$, since the inputs to the repetition node are $\calN(2\mu,4\mu)$ and $\calN(0,4\mu)$. The distributions that are not a result of~\eqref{eq:subtraction} but are still altered are in blue.}
    \end{subfigure}
    \caption{Illustration of \ac{CDEGA} for an error in $N=4$.}
\label{fig:CDEGAFig}
\end{figure}

If a decoding error is made, an erroneous $1$ now exists in the all-zero codeword. The effect of this erroneous bit on posterior \acp{LLR} for future bits is as follows. Note that the decoding tree only contains degree-two repetition nodes and parity-check nodes. As a result, the recursive functions of $\mu$ at many future repetition nodes (at all layers) will now update to
\begin{align}
\mu_{i}^{k} &= \mu_{i}^{k+1} - \mu_{i-2^{k}}^{k+1},\label{eq:subtraction}
\end{align}
for any bit $u_{i-2^{k}}^{k+1}$ that gets updated to a $1$. We are now \emph{subtracting} the two positive mean inputs. Up until this point, because of the structure of the problem, both $\mu_{i}^{k+1}$ and $\mu_{i-2^{k}}^{k+1}$ characterize identical densities $\mathcal{N}(\mu_i^{k+1}, 2\mu_i^{k+1})$. Hence, the new distribution resulting from ~\eqref{eq:subtraction} is of the form $\mathcal{N}(0, 4\mu_i^{k+1})$. Because polar codes have a tree structure the output of a check node is always smaller than either of the two inputs, the expectations of all future \acp{LLR} shrink after the first error, with many becoming zero. An example of this occurrence is illustrated in Figure \ref{fig:CDEGAFig}. We acknowledge that \cite{fengTailBitingPolarizationAdjustedConvolutional2023} identified a similar phenomenon, but did not leverage it or explicitly show what happened to future \acp{LLR}.

Since the new distributions are no longer symmetric nor identical, the equations corresponding to repetition and check nodes now need four inputs (two means and two variances), as opposed to just $\mu_i^k$. Repetition nodes still remain as the sum or difference of the two input distributions. However, calculating the propagation of distributions through a check node becomes more challenging than traditional DEGA because the $\phi$ function is no longer appropriate. Empirical methods, such as Monte Carlo or Gauss–Hermite quadrature are instead implemented to estimate the output distribution of check nodes, calculating both mean and variance. Even in the absence of symmetry, where small, if not negative, posteriors may have a large variance, we still assume that all inputs and outputs of the check nodes and repetition nodes are Gaussian. If the conditional posterior of bit $i$ comes from a Gaussian distribution $\calN(\mu_i,\sigma^2_i)$, the expected \ac{BM} can then be calculated with one of the two integrals 
\begin{align}
    \mathbf{E}[\textrm{BM}_i] = \begin{cases}
                    \int_{0}^{+\infty}|\lambda|\mathcal{N}(\mu_i,\sigma^2_i)   \\
                    \int_{-\infty}^{0}|\lambda|\mathcal{N}(\mu_i,\sigma^2_i) 
    \end{cases},
\end{align}
depending on the optimal decision for $u_i$ based on the sign of $\mu_i$. The decision of $u_i^0$ for frozen bits is fixed. Because different formulas exists for calculating the BM in~\eqref{BM} and the check nodes required in~\eqref{info} (e.g., min-sum or modified min-sum), the exact equations presented may also differ, however, the ideas presented remain consistent. Since the PM is a sum of all branch metrics, it can be estimated as 
\begin{align}
    \mathbf{E}[\textrm{PM}] = \sum_{i=0}^{N-1}\mathbf{E}[\textrm{BM}_i], \label{totalPM}
\end{align}
 %This means that frozen bits with conditionally large variances and small means of the posterior LLR contribute greatly to the \ac{PM} since their $\mathbf{E}[\textrm{BM}]$ are large. 
 where $\mathbf{E}[\textrm{PM}]$ is the expected \ac{PM} of any specific path averaged over all noisy realizations of the all-zero codeword. For information bits, since the decoder makes decisions based on the posterior \ac{LLR} and \ac{BM}, one of the two options to branch $u_i^0$ has no penalty, meaning~\eqref{BM} assigns no increase in the \ac{PM}. We refer to branch decisions such as these as greedy. This means that much of the contribution to  $\mathbf{E}[\textrm{PM}]$ comes from the frozen bits with a large $\mathbf{E}[\textrm{BM}]$. With a traditional polar code, this arises when the realization of the \ac{LLR} is negative, either from a small enough mean or large enough variance. The contribution of a precoder comes from a forced decision of $1$ on a frozen bit with a large conditional mean. 

%which are ones with a small conditional mean but large variance, or ones with a large conditional mean and a forced decision of $1$ as a result of the convolutional precoding.  TODO: what was i trying to say here 

Figure \ref{fig:CDEGAFig} illustrates a scenario in which all bits up to $i$ have been decoded but an error was made on the $i$th bit. \ac{DEGA} becomes inaccurate in estimating the posteriors of the future bits. Thus, computing the conditional \ac{DEGA} distributions of future bits is essential for estimating the PM of this error path. 

\begin{figure}[ht]
    \centering

    % Top-left
        \centering
        \includegraphics[width=\linewidth]{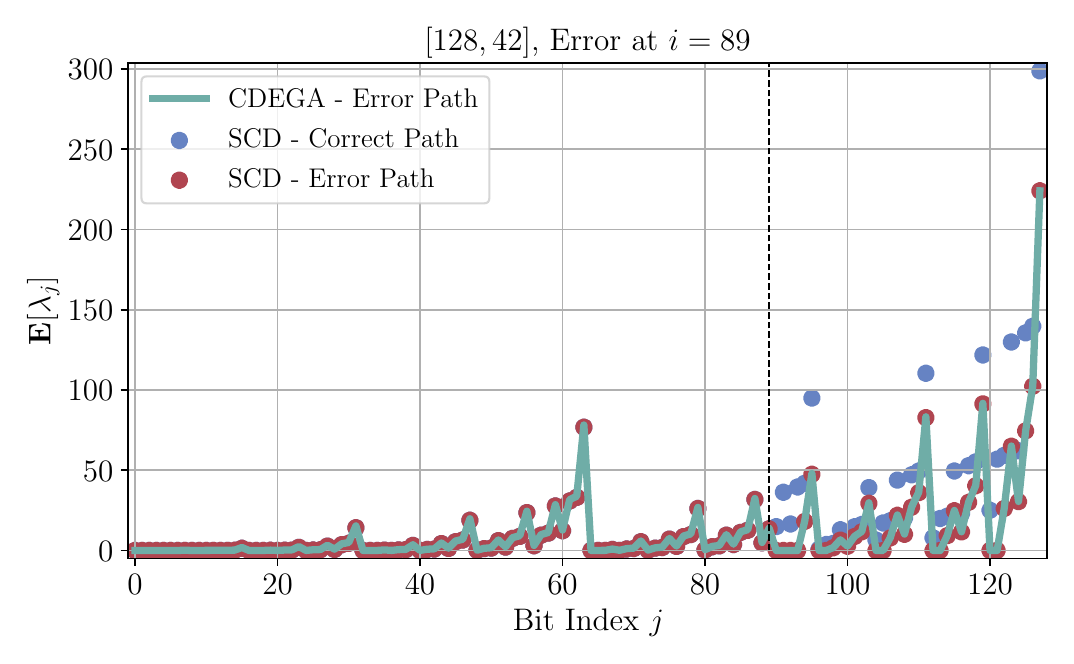}
        \captionof{figure}{The mean of the posterior \acp{LLR} when decoding a $[128,42]$ polar code. Two paths are shown via SCD: one of the correct, all-zero path; one erroneous path where exactly one error occurs on $i=89$. Plotted alongside is the CDEGA estimation on the same erroneous path, which is able to accurately track the mean.}
        \label{fig:mu-track}
    \hfill
\end{figure}

\begin{figure}[ht]
    \centering

    % Top-left
        \centering
        \includegraphics[width=\linewidth]{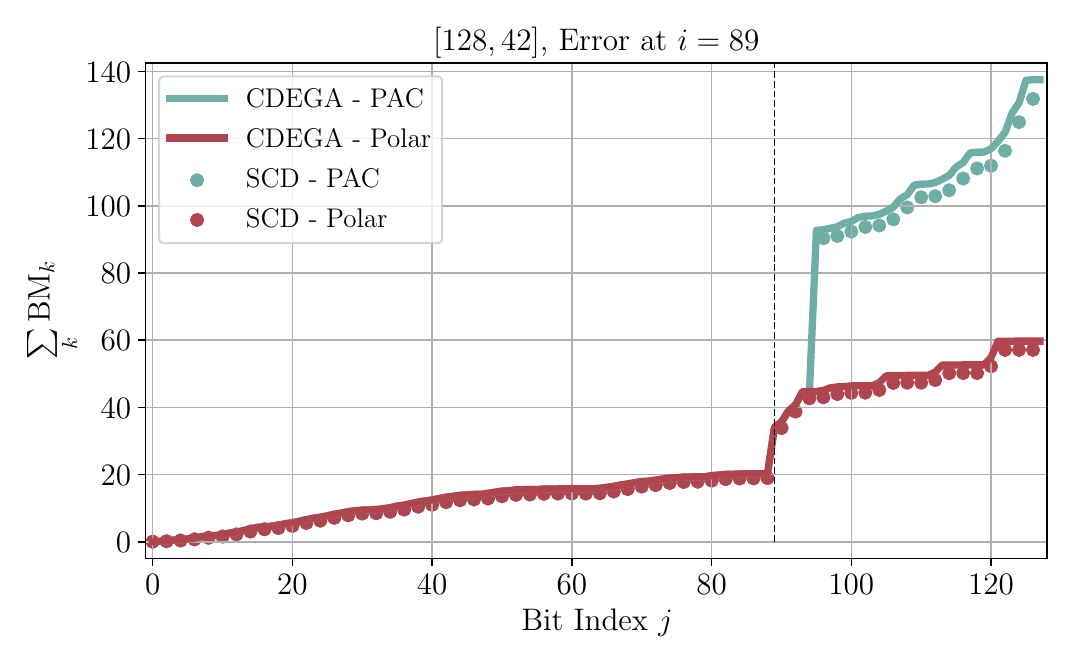}
        \captionof{figure}{The average addition of the \acp{BM} when decoding a $[128,42]$  code with an error on bit $i = 89$. The CDEGA prediction of the BM is shown alongside the real SCD, for both polar and PAC codes.}
        \label{fig:scd-track}
    \hfill
\end{figure}

We show, in Figures \ref{fig:mu-track} and \ref{fig:scd-track}, an example of CDEGA for \mbox{$N = 128, K = 42$} with the RM-Polar construction \cite{RMPolar}. The all-zero codeword is transmitted, and \emph{only} bit $i = 89$ is forced to be an erroneous $1$, whereas the rest of the codeword is assumed to be $0$. This allows us to visualize how \textit{exactly one} error is tracked. Since we know exactly what error is made, we only need SCD to examine the path. Figure \ref{fig:mu-track} depicts \acp{LLR} obtained through Monte Carlo trials of SCD on the correct, all-zero path and error path, together with the \ac{CDEGA} estimation of the \acp{LLR} of the error path. \ac{CDEGA} accurately estimates the \acp{LLR} of the erroneous path. Additionally, CDEGA accurately captures how the \acp{LLR} immediately collapse after the error and remain less than the correct \acp{LLR} for all future bits. Figure \ref{fig:scd-track} illustrates that by using estimations of the \acp{LLR}, \ac{CDEGA} accurately tracks the \acp{BM}, and ultimately, the \ac{PM}. The approach can be adapted to account for a precoder in a \ac{PAC} code, also seen in \mbox{Figure \ref{fig:scd-track}}. \ac{CDEGA} accurately tracks the magnitudes  of the \acp{BM}, allowing a bit-level analysis of how the PM updates on any decoding path.

\subsection{CDEGA Algorithm}
We now develop a first algorithm to construct information sets based upon CDEGA, detailed in Algorithm \ref{alg:algorithmCDEGA}. The algorithm calculates the mean of the posterior \acp{LLR}, $\mu_i^0$, according to~\eqref{eq:checknode},~\eqref{eq:repitition}, and a design \ac{SNR}. The larger $\mu_i^0$ is, the more reliable bit $u_i^0$ is. The algorithm initializes $\mathcal{A}$ to the \ac{DEGA} rate profile, which consist of the $K$ most reliable bits. It also selects the next $\epsilon$ most reliable bits, which we denote as $\mathcal{F}' \subseteq \calF$. Then \ac{CDEGA} is used to estimate the \ac{PM} on the path that incorrectly branches at bit $i$. As a first-order accounting of the precoder $T$, the frozen bits reached from $i$ through $T$ are flipped, but the effect of other information bits on frozen bits are ignored. This calculation is $\mathrm{PM}(i,\mathcal{A},T)$ in Algorithm \ref{alg:algorithmCDEGA}. Then, the bit with the \textit{smallest} PM is taken from $\mathcal{A}$, and frozen to $\mathcal{F}'$. $\mathrm{PM}(i,\mathcal{A},T)$ is applied on the bits in $\mathcal{F}'$ and the bit with the \textit{largest} PM is added to $\mathcal{A}$. This is repeated until we have created an information set already visited, or until a maximum number of iterations is reached. The goal of this algorithm is estimating \[
\max_{\mathcal{A}} \min_{i \in \mathcal{A}} \mathrm{PM}_i, 
\]
which aims to find the information set $\mathcal{A}^*$ such that any single error has a large PM, allowing the decoder to more easily determine the correct path. As seen in Figures \ref{fig:64_32_FER}-\ref{fig:512_256_FER} the FER of the resulting constructions is competitive, giving nearly a dB improvement over \ac{DEGA} for $N=256$. Additionally, the complexity of evaluating $\mathrm{PM}(i,\mathcal{A},T)$ on any individual path is approximately $\mathcal{O}(M N\log N)$  when using Monte Carlo with $M$ trials, or  $\mathcal{O}(Q^2 N\log N)$ when using Gauss-Hermite quadrature with $Q$ points.

\begin{algorithm}[ht]
\caption{CDEGA Information-Set Search}
\label{alg:algorithmCDEGA}
\KwIn{$N,K$, construction SNR, precoder $T$, candidate size $\epsilon$}
\KwOut{Information set $\mathcal{A}^*$}

Initialize $\mathcal{A}$ as the $K$ most reliable DEGA channels\;
Let $\mathcal{F}'$ be the next $\epsilon$ most reliable frozen channels\;
$\mathcal{A}^* \gets \mathcal{A}$\;
$P^* \gets \min_{i\in\mathcal{A}} \mathrm{PM}(i,\mathcal{A},T)$\;

\While{$\mathcal{A}$ has not been previously visited}{
    $i^* \gets \arg\min_{i\in\mathcal{A}} \mathrm{PM}(i,\mathcal{A},T)$\;
    $\mathcal{A}' \gets \mathcal{A}\setminus\{i^*\}$\;
    $\mathcal{F}' \gets \mathcal{F}'\cup\{i^*\}$\;

    $j^* \gets \arg\max_{j\in\mathcal{F}'}
    \mathrm{PM}(j,\mathcal{A}'\cup\{j\},T)$\;

    $\mathcal{A} \gets \mathcal{A}'\cup\{j^*\}$\;
    $\mathcal{F}' \gets \mathcal{F}'\setminus\{j^*\}$\;

    $P \gets \min_{i\in\mathcal{A}} \mathrm{PM}(i,\mathcal{A},T)$\;
    \If{$P>P^*$}{
        $\mathcal{A}^* \gets \mathcal{A}$; \quad $P^* \gets P$\;
    }
}
\Return{$\mathcal{A}^*$}\;
\end{algorithm}

\subsection{Relation between conditional DEGA and code weight}
\label{sec:PMWeight}
As seen in Figures \ref{fig:CDEGAFig} and \ref{fig:mu-track}, some of the conditional posteriors after an error have zero mean. Decisions on these bits do not alter the expected \ac{PM}. Additionally, the decisions on such bits do not alter the magnitude of the future expected \acp{LLR}, although they may change the sign. In other words, neither decision is viewed as incorrect from the perspective of the decoder. These bits are problematic since they have equal probability to be $1$ or $0$. These bits interfere with the precoding, either by creating even parity and not altering a frozen bit decision, or by swapping the sign of a frozen bit that will be altered by \ac{PAC}. Both scenarios limit the potential increase in \ac{PM} incurred by \ac{PAC} and create paths that have less than expected \acp{PM}. 

We compare the finding of zero-mean posteriors to  \cite{rowshanFormationMinweightCodewords2023}, in which Rowshan et al. define the set 
\begin{align}
     \mathcal {K}_{i}\triangleq \{j \in [i+1,N-1]\colon \mathop {\mathrm {w}}\nolimits (\mathbf {g}_{j})\geq \mathop {\mathrm {w}}\nolimits (\mathbf {g}_{i}+ \mathbf {g}_{j})= \mathop {\mathrm {w}}\nolimits (\mathbf {g}_{i})\},
\end{align}
where ${\mathrm {w}}(\cdot)$ is the Hamming weight, and $\mathbf {g}_{j}$ is row $j$ in the $G^{\otimes n}$ matrix. In words,  $\mathcal {K}_{i}$ is the subset of all row vectors of $G^{\otimes n}$ with a weight greater than or equal to $\mathrm{w}(\mathbf{g}_i$), but linearly combine with $\mathbf {g}_{i}$ in a way such that the resulting Hamming weight is the same as $\mathrm{w}(\mathbf {g}_{i})$. In other words, if row $\mathbf{g_i}$ is the codeword, no inclusion of an additional row from $\mathcal{K}_i$ will affect the codeword weight. Rowshan et al. use these sets to help with min-weight codewords enumeration.  One can also focus on rows with smaller row weights than $\mathrm{w}(\mathbf {g}_{i})$, and define a similar set 
\begin{align}
     \mathcal {K}^{*}_{i}\triangleq \{j \in [i+1,N-1]\colon \mathop {\mathrm {w}}\nolimits (\mathbf {g}_{j}) < \mathop {\mathrm {w}}\nolimits (\mathbf {g}_{i}+ \mathbf {g}_{j})= \mathop {\mathrm {w}}\nolimits (\mathbf {g}_{i})\}.
\end{align}

Empirically, $\mathcal{K}_i \cup \mathcal{K}^*_{i}$ seems to correspond to all the bit positions that have a zero-mean posterior conditioned on an error at bit $i$. 
In other words, deciding $\hat{u}_j$ incorrectly once an error has been made on $\hat{u}_i$ neither affects the resulting PM \emph{nor} codeword weight. This suggests a conceptual connection between our work on \ac{CDEGA} and \cite{rowshanFormationMinweightCodewords2023}. This does not mean, however, that two codewords with the same weight  have the same PM; the PM might therefore offer more information. Additionally, the PM is directly tied to list decoding, whereas other metrics, including weight are not. Thus, the PM might be a more accurate metric to leverage than weight. Regardless, since weight appears to be a closely tied proxy to PM, the work in \cite{rowshanFormationMinweightCodewords2023} could be used as a valuable tool to creating sets aligning with our design.

\subsection{Union bound with non-ML decoders}

Any linear block code can be examined through its weight spectrum $\{A_d\}_{d=0}^N$, where $A_d$ represents the number of codewords with weight $d$, with $\sum_{d=0}^{N} A_d = 2^K$ for an $[N,K]$ code.

Given the weight spectrum, the frame error rate under an ML decoder over the AWGN channel can be bounded by the union bound, defined as: 
\begin{align}
    \label{eq:unionbound} 
    \textrm{FER} \leq P_{\mathrm{UB}} = \sum_{d = d_{\min}}^{N} A_d Q\big(\sqrt{2dRE_b/N_0}\big), 
\end{align}
where $d_{\min}$ is the minimum distance, $R$ is the rate of the code, $E_b/N_0$ is the \ac{SNR}, and $Q$ is the Q-function. 

Furthermore, at high SNRs, the union bound is dominated by low weight codewords, specifically $d_{\min}$ and thus can be approximated by 
\begin{align}
    P_{\mathrm{UB}} \approx A_{d_{\min}} Q\big(\sqrt{2d_{\min}RE_b/N_0}\big).
\end{align}
Consequently, a large portion of the literature is focused on finding codes that improve  $P_\mathrm{UB}$, which is satisfied by either increasing $d_{\min}$ or decreasing $A_{d_{\min}}$.

\begin{figure*}[ht] 
    \centering 
        \begin{subfigure}[t]{0.48\textwidth} 
            \centering \includegraphics[width=\textwidth]{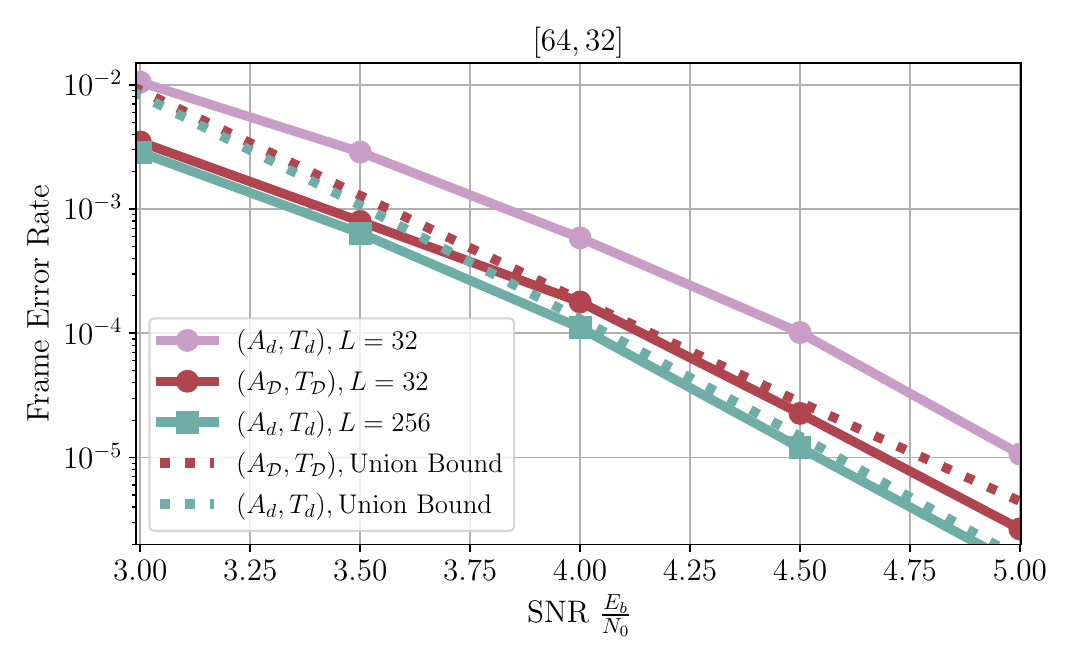} 
            %\caption{FER vs SNR, $L = 32, 256$}
            \label{fig:dmin256} 
            \end{subfigure}
            \hfill 
            \begin{subfigure}[t]{0.48\textwidth}
            \centering 
            \includegraphics[width=\textwidth]{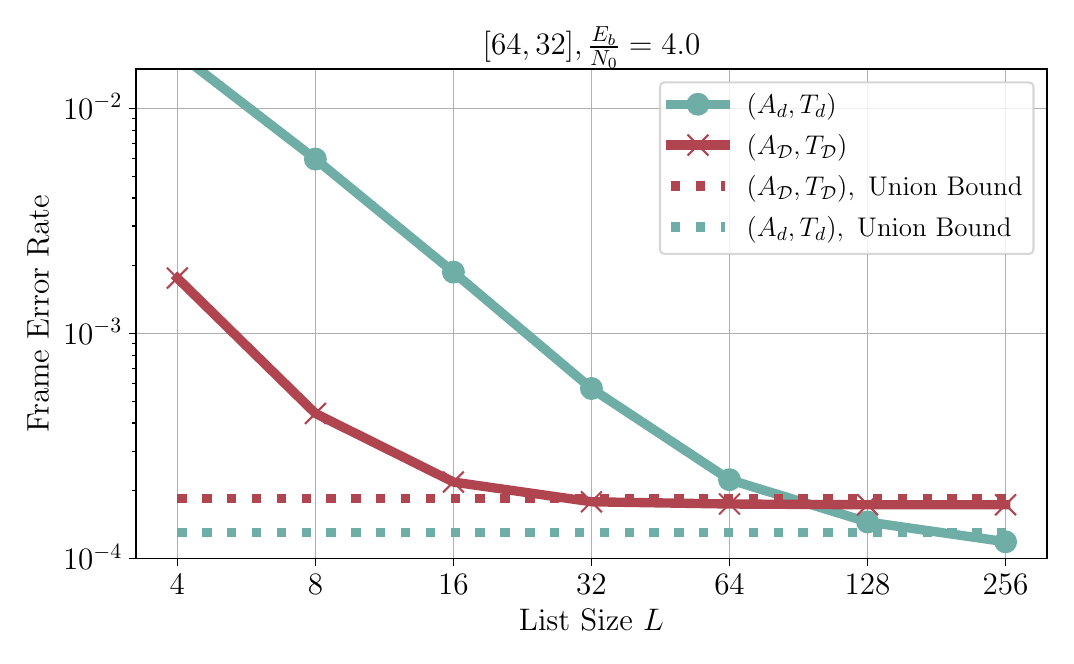} %\caption{FER vs L, $\Eb$}
            \label{fig:dmin32} 
            \end{subfigure} 
        \vspace{-5mm}
        \caption{A comparison between two precoded polar codes for $N=64,K=32$ with two constructions: one from our algorithm, $(\mathcal{A}_\calD,T_\calD)$, and one designed to optimize $A_d$, $(\mathcal{A}_d,T_d)$. Although $(\mathcal{A}_d, T_d)$ provides a better $P_{\mathrm{UB}}$, the finite-list performance can differ substantially. At $L=32$, $(\mathcal{A}_d,T_d)$ performs significantly worse than $(\mathcal{A}_\calD,T_\calD)$. The performance does not improve over the $\calD$-construction unless decoded with a very large list size. Therefore, at practical $L$,  weight spectrum alone may not be a sufficient condition in determining the best codes.} \label{fig:dmin} \end{figure*}

While this bound holds for an ML decoder, it is often assumed that a polar code with a sufficient list size approximates the performance of ML. While this is true for a large enough list size, a practical size, e.g., $L =32$, may not be sufficient. This means that one code may empirically outperform another, despite having a larger $P_\mathrm{UB}$. To demonstrate this, we compare two precoded polar codes in Figure \ref{fig:dmin}. One of these two codes, denoted as $(\mathcal{A}_\calD,T_\calD)$ is the code we construct using algorithms explained in future sections. The specifics of the algorithm are not necessary for this comparison, suffice to say that we are optimizing metrics that are tied to SCL decoder at a specific $L$. The second code, denoted as $(\mathcal{A}_d,T_d)$ was found over a brute-forced search for codes with large minimum distance. Both constructions change not only  $\mathcal{A}$, but also the precoder $T$, creating non-standard, time-varying precoders. The minimum distance for $(\mathcal{A}_d,T_d)$ is $10$, whereas it is only $8$ for $(\mathcal{A}_\calD,T_\calD)$.  Additionally, as seen in Figure~\ref{fig:dmin}, when enumerating both spectra in their entirety, $P_{\mathrm{UB}_d}$ is lower than $P_{\mathrm{UB}_\calD}$, implying better performance. For a very large list size, $L \geq 128$, the bound is an accurate estimator of performance, but for a list size of $L = 32$, $(\mathcal{A}_\calD, T_\calD)$ outperforms  $(\mathcal{A}_d, T_d)$. Therefore, while the performance of SCL at large list sizes is well approximated by~\eqref{eq:unionbound}, at smaller list sizes one must consider additional metrics directly tied to SCL. It is worth noting that the weight spectrum of $(\mathcal{A}_\calD,T_\calD)$, as well as the spectra of the other codes we present, do substantially improve over standard \ac{PAC} codes.

\section{Tail Events} 
\label{sec:tail-events}

\begin{figure}[ht]
    \centering
        \centering
        \includegraphics[width=\linewidth]{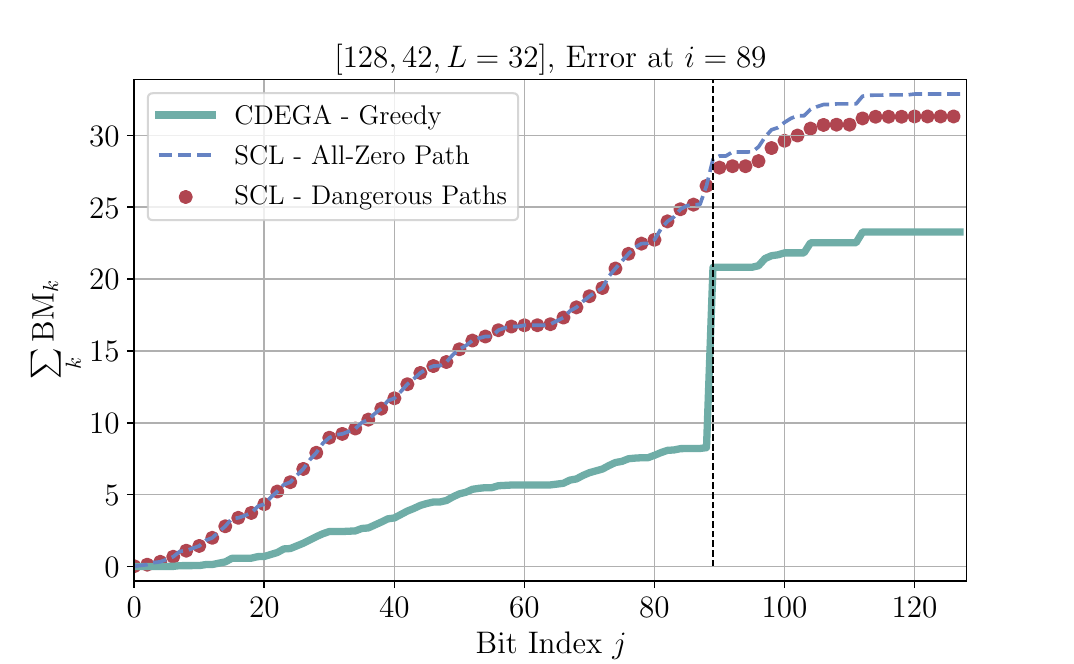}
        \captionof{figure}{The average addition of the \acp{BM} when decoding a $[128,42, L = 32]$ polar code, where the dangerous path diverged at bit $i=89$. The CDEGA prediction of the BM is shown alongside the lowest-PM path and all-zero path of erroneous trials from a decoding simulation.}
        \label{fig:scl-track}
    \hfill
\end{figure}

The metrics $\mathbf{E}[\lambda_i]$ computed by \ac{CDEGA} are expectations over all noise realizations. However, average metrics do not fully capture error events that occur in a decoder on an abnormal realization of the channel, which we refer to as a tail event. 
To illustrate this difference, Figure \ref{fig:scl-track} again tracks the \acp{BM} for the same code, but using \ac{SCL} with $L = 32$. Importantly, instead of averaging over all trials, we average over \textit{only} the trials with an error, and look at the MMPM on these erroneous trials. We again refer to erroneous paths that beat the correct path as dangerous paths. By only averaging over the erroneous trials, as opposed to all trials, the metrics more accurately capture what occurs on a tail event. The BM accumulation in Figure \ref{fig:scl-track} is far different from that in Figure \ref{fig:scd-track}. Notably, the BMs \emph{before} the error are much higher, indicating the poor quality of the channel realization that creates an error. Additionally, the actual error occurring on bit $i = 89$ induces negligible BM, which makes it difficult for the decoder to disambiguate this error. Lastly, nearly no BM accumulates on the information bits after \mbox{bit $i$}, meaning that the dangerous path is composed of greedy branch decisions, as postulated in \mbox{Section \ref{sec:cdega}}. To account for this, the \ac{CDEGA} estimation in Figure \ref{fig:scl-track} also makes greedy decisions on information bits, setting bit $i$ for $i \in \mathcal{A}$ and $ i > 89$ to the value that  has $\textrm{BM}_i = 0$. Interestingly, the shape and scale of the BMs \emph{after} the error are tracked with decent accuracy, showing some consistency. However, the BMs before and up to bit $i$ are difficult to account for, as the standard Gaussian assumption fails on these tail events. Ultimately, while \ac{CDEGA} can broadly analyze what occurs on an erroneous decision, and Algorithm \ref{alg:algorithmCDEGA} remains somewhat robust, focusing specifically on the tail events and dangerous paths allows for an increase in performance.

\subsection{Dangerous paths}

Analyzing dangerous paths is challenging, as there is no obvious correlation between the tail events and the errors that they cause. Therefore, we pivot to a sample-based approach, with our analysis coming from Monte Carlo that simulates decoding the all-zero codeword over $M$ realizations of a channel. Whereas in the previous section, we analyzed the error path unconditionally, we will now analyze only those still remaining in the final list 
$\mathcal{L}^*_m$ of a Monte Carlo trial $m$.

Under this sample based framework, we will be looking at empirical data. We denote the all-zero correct path as $\ell_0$. For a surviving, incorrect path $\ell$ in $\mathcal{L}_m^* \setminus \{\ell_0\}$, the \emph{path gap} on trial $m$ is
\[
  D_{\ell,m} \;=\; \mathrm{PM}_m(\ell \phantom{}) \,-\, \mathrm{PM}_m(\ell_0),
\]
and the \textit{decision gap} is
\begin{equation}
  D_m \;\triangleq\; \min_{\ell \in\calL^*_m \setminus \{\ell_0\}} D_{\ell ,m}.
\end{equation}

The decoder makes a correction decision if and only if $D_m > 0$ and $\ell_0 \in \mathcal{L}_m^*$. A decoder that fails the first condition is a selection error, and a decoder that fails the second condition is a path-loss error. We call a Monte~Carlo trial
\emph{dangerous} if $D_m \leq 0$. A dangerous path is a path that produces $D_{\ell, m} \leq 0$, and a dangerous path arises on a tail event. The set of dangerous paths is denoted as $\calL_m^- \subseteq \calL_m^*$, and we denote any individual dangerous path as $\tilde{\ell}$.

\subsection{Dangerous-path spectrum}

With consistent notation, we define the information vector of dangerous paths as $\tilde{v}$ and  
$\tilde{u} = \tilde{v}T$. Subscript $(j,\ell,m)$ denotes bit $j$ on path $\ell$ at Monte Carlo trial $m$. For each position \mbox{$j \in \{0,\dots,N-1\}$}, we define

\begin{align}
    \delta_j = \frac{1}{M}\sum_{m=1}^{M}\sum_{\calL^-_m}
      \mathbf{1}\{\tilde v_{j,\tilde{\ell},m}=1\}.
\end{align}

We also define a similar metric for bits coming \emph{after} precoding, namely

\begin{align}
    \rho_j = \frac{1}{M}\sum_{m=1}^{M}\sum_{\calL^-_m}\mathbf{1}\{\tilde u_{j,\tilde{\ell},m}=1\}.
\end{align}

In words, a large $\delta_j$ means that $v_j$ is consistently decoded incorrectly on a dangerous path.  This includes both the situation in which it is the first error \emph{and} when it is an additional error on an already erroneous path. Note that $\delta$ is looking at $v$ and \emph{not} $u$. This means that for $j \in \mathcal{F}$, $\delta_j = 0$, and thus only information bits are relevant with regards to $\delta$. Regardless of whether it is causing the dangerous path or abetting it, if bit $j$ has a large $\delta_j$, then it is a large contributor of the error rate and likely should be frozen. It is worth noting that it is actually $u_j$ that interacts with $\textrm{BM}_j$ and not $v_j$. Therefore, it may seem natural to look at $\rho_j$ instead of $\delta_j$. Suppose, a bit were to have a small $\delta_j$ but a large $\rho_j$. While this might more strongly suggest involvement with the \ac{PM}, it also means that freezing the bit would change little, since $v_j$ is typically $0$ already. Therefore the driving metric of \emph{bad} bits that should be frozen (removed from $\mathcal{A}$) will be those with the highest $\delta_j$. When referring to metrics without regard to a specific bit index, we drop the subscript, e.g., just$~\delta$. We call the $\delta$-based set design dangerous-path occupancy.   

While $\delta$ is useful for determining which bit to freeze, $\rho$ provides insight for already frozen bits. Any $j$ (in  $\mathcal{A}$ or $\mathcal{F}$) may have a non-zero $\rho_j$. For a frozen bit $j \in \mathcal{F}$, a large $\rho_j$ means that bit $j$ is currently being utilized under the precoder. Therefore, even if $j$ is a promising candidate to add to $\mathcal{A}$, unfreezing it will interfere with previous bits utilizing it, and inclusion in $\mathcal{A}$ it may come at a performance cost elsewhere.   

All tail-event metrics captured from real SCL decoding, including $\delta$, $\rho$, $\lambda$, are collectively labeled as $\mathcal{D}$, the dangerous-path spectrum. Importantly, $\calD$ is calculated from \emph{all} dangerous paths where $D \leq 0$, not just the one path selected by SCL. This means that path-loss errors influence $\calD$ more as one path-loss error can add up to $L$ data points. In order to calculate $D$ on path-loss errors, the all-zero path is additionally decoded with a decoder that forces the all-zero path to be selected, ensuring all trials have the all-zero $\mathrm{PM}$. This means that not all $L$ paths are necessarily added even in a path-loss error.

\subsection{Information set design}
We now provide a high level overview of how we construct information sets. We initialize an information set $\mathcal{A}$ of our $[N,K,L]$ code with a standard profiling tool, e.g., DEGA. We then simulate list decoding of the current PAC code to obtain $\mathcal{D}$. We then remove the information bit with the largest $\delta$. When scoring frozen bits, we first threshold based on a percentile of $\rho$ (in our trials, $50^{\mathrm{th}}$ percentile), meaning that those with a large $\rho$ automatically get rejected before any additional screening. The frozen bits that make it past the threshold are scored by the $10^{\mathrm{th}}$ percentile of the LLR of bit $j$, $q_{.1}(|\lambda|_j)$. A large $10^{\mathrm{th}}$ percentile means that even in lower tail events of reliability, the decoder will still make a reliable decision on $u^0_j$, if it were to become an information bit. The most reliable frozen bit is selected to be added to $\mathcal{A}$. These heuristics ensure that a frozen bit selected will perform well (because of the reliability) without lowering the path metrics of other incorrect paths (because of the small $\rho)$.

Removing one information bit while simultaneously another ensures that the new set will still have $K$ information bits, and this process of swapping bits between $\mathcal{A}$ and $\mathcal{F}$ is continued for a set number of iterations (an algorithmic parameter) or until the empirical \ac{FER} does not decrease. 

Note that we have looked at two heuristics for determining the least reliable information bits: CDEGA and dangerous-path occupancy. CDEGA aims to estimate the information bit where the most error paths stem from; dangerous-path occupancy aims to estimate the information bit that occurs the most in error paths, regardless of where the original error bit was. Freezing the bit with the largest CDEGA metric will, by design, remove the possibility of that bit being the first erroneous path, but in practice, we find that it rarely affects other paths. However, empirically we find that removing the bit with the largest $\delta$ affects numerous paths, leading to an improvement in \ac{FER}. This leads us to the claim that \ac{CDEGA} focuses on how reliability evolves after a wrong branch survives, whereas dangerous-path occupancy focuses on which rate-profile intervention can suppress many surviving competitors at once. From this perspective, as well as empirically, algorithmic decisions based on $\delta$ seem to perform better.

\subsection{Precoder design}

For high-performing information sets and generator connections, the precoder alters the decision of some frozen bits with non-zero conditional posterior, thus increasing the PM of the incorrect path. Note that for an information bit, the decoder always has the choice of both $\{0,1\}$, so an \ac{SCL} decoder most likely favors the choice that does not increase the path metric. Therefore, the ability to alter the decision of \textit{frozen} bits might be the underlying reason of PAC code's success, and additionally why this success may be dependent on decoders that can track multiple paths. 

To demonstrate this hypothesis, we propose an additional algorithm, detailed in Algorithm~\ref{alg:precoder_design}. The information set used in Algorithm~\ref{alg:precoder_design} is obtained from Algorithm~\ref{alg:set_design}, which was tailored to a generic, time-invariant precoder, $T_0$. However, the initial precoder of Algorithm \ref{alg:precoder_design} is the identity matrix $I$.  We then simulate list decoding of the current PAC code and, for
every dangerous path $\tilde{\ell} \in\calL^-_m$ of every trial $m$, we compute the effect of each
candidate edge \mbox{$(i,j)$, $i\in\mathcal A$, $j\in\mathcal F$, $i<j$} according to:
\begin{align}
\Delta_{\tilde{\ell},m}(i,j) =
\begin{cases}
(1-2\tilde u_{j,\tilde{\ell},m})\,\lambda_{j,\tilde{\ell},m} & \textrm{if}\quad \tilde v_{i,\tilde{\ell},m} = 1 \\
0 & \textrm{if}\quad \tilde v_{i,\tilde{\ell},m} = 0
\end{cases},
\end{align}

In other words, $\Delta_{\tilde{\ell},m}(i,j)$ is an estimate of the new path gap increase if a precoder edge were placed between $i$ and $j$, and $\mathrm{BM}_{j,\tilde{\ell},m}$ were altered based on the decision of $i$ and its own conditional posterior. Each iteration keeps a score, $S(i,j)$, for every pairing, which is the number of \emph{saved} trials where \mbox{$D_{\tilde{\ell},m} + \Delta_{\tilde{\ell},m}(i,j) > 0 \; \forall \tilde{\ell} \in \calL_m^-$}. The edges with the largest score are then further evaluated, and the edge that most improves the empirical FER is permanently added. Note that by looking at $1-2\tilde{u}_{j,\tilde{\ell},m}$, we are ensuring that the edge added  prioritizes bits where the decision of $\tilde{u}_{j,\tilde{\ell},m}$ matches the sign of $\lambda_{j,\tilde{\ell},m}$, since adding edges where $1-2\tilde{u}_{j,\tilde{\ell},m}$ and $\lambda_{j,\tilde{\ell},m}$ \textit{already} disagree would actually further decrease the PM (see Section \ref{sec:cdega}). Importantly, $S(i,j)$ increments only if \textit{all} dangerous paths are corrected, as an error would still occur if only some paths were corrected. This heuristic creates precoders that: 
\begin{enumerate}
    \item Vary in time and length, 
    \item Only connect information bits to frozen bits.
\end{enumerate}

Algorithmically, we allow for $k \geq 1$ edges to be added per iteration. Empirically, we find that a slightly larger $k$ \mbox{(e.g., $2$)} seems to converge faster and to a better FER than $k=1$.  An additional algorithmic constraint, that is not seen in ML decoding, is what we call $\ell_0$-aware weighting. Specifically, we separate the path-loss and selection errors. If it is a selection error, $j$ can be any frozen bit after $i$. If it is a path-loss error, we fix \mbox{$\Delta_{\tilde{\ell},m}(i,j) = 0$ if $j > e_{\tilde{\ell},m} + b$}, where $e_{\tilde{\ell},m}$ is the first error on path $\tilde{\ell}$ and $b$ is an algorithmic parameter. The idea is that if it is a path-loss error in which $\ell_0$ does not survive until the end, there is urgency to correct before pruning, which is not important in a selection error. The parameter $b$ is a proxy for this urgency, with the goal being finding edges that, even if not optimal, change the PMs of erroneous paths enough to avoid losing $\ell_0$ when pruning from $2L$ to $L$ paths. We choose $b = 32$, and conjecture that $b$ should grow with list size.  

\subsection{Additional algorithmic decisions}

We provide full pseudocode of the two algorithms below, as well as provide the code and sets on GitHub \cite{KannGithub}. In addition, we highlight some of the salient features in \mbox{Algorithms \ref{alg:set_design} and \ref{alg:precoder_design}}. Within both algorithms, the optimization objective of frame error rate is obtained via Monte Carlo simulations of an SCL decoder. While Monte Carlo is costly, most estimation functions of code performance, including detailed weight enumeration, are generally of nontrivial complexity. Here, an advantage the Monte Carlo provides is an ability to directly optimize an empirical estimate of the frame error rate, as opposed to a proxy metric. Additionally, the Monte Carlo simulation fully accounts for the list size and the nature of the channel, in our case, AWGN. 

Two Monte Carlo simulations are used in Algorithms \ref{alg:set_design} and \ref{alg:precoder_design}, FullMC and MiniMC. As the name suggests, the FullMC is the full scale Monte Carlo evaluation, which calculates the \ac{FER}, as well as the dangerous-path spectrum, $\mathcal{D}$. This is done by running many trials of an SCL decoder, which in our work is typically $4\times 10^5$ trials. The MiniMC, used in Algorithm \ref{alg:set_design}, also samples trials of an SCL decoder, except here it is typically only $1\times10^5$ trials. The idea is to use the MiniMC to find a rough estimate on the FER, and then only use the FullMC on promising candidates, avoiding excess simulations on non-competitive candidates.

In Algorithm \ref{alg:precoder_design}, $\Delta_{j,\tilde{\ell},m}$ alone may itself not be sufficient, as an edge change that removed the dangerous paths on one trial may create dangerous paths on a different trial. Therefore, when calculating FullMC, we keep track of the noise realizations that caused errors \textit{and} near misses, which are successful decoding trials where $0 < D_m \leq \tau$, where $\tau$ is an algorithmic parameter. Then, candidates with a large $S(i,j)$ are only tested on these specific noise realizations.  The function in Algorithm \ref{alg:precoder_design} to test the new edges of the precoder matrix is called Replay, and the trials captured are referred to as the replay trials, $\calR$. The idea is that by only using these trials, the screening is much faster and avoids fresh-sampling variance by resampling on the same realizations. Although no choice of precoder can alter the PM of the all-zero path, a trial that is not in $\calR$ can become erroneous if the new edge lowers the PM of some incorrect path $\ell$ by some value $\alpha > D_{l,m} > \tau$. Therefore a larger $\tau$ lowers the approximation error at the cost of more trials used. The edges that get past this screening are then finally evaluated with FullMC. The Monte Carlo trials for the current set, candidates, and the numerical evaluations in Section~\ref{sec:results} are all from separate seeds at the fixed design SNR.

We emphasize that neither algorithm performs an exhaustive search. Instead, each iteration runs a single Monte Carlo simulation of the
\emph{current} code, and the resulting dangerous-path spectrum $\calD$ is used to
deterministically rank a small number of local moves. Thus Monte Carlo
is used to \mbox{(i) measure} $\calD$ for the current code and (ii) verify a few proposed
moves, while the choice of \emph{which} moves to try is dictated by $\calD$. The complexity of the algorithm is dominated by the Monte Carlo simulation.

\section{Results}
\label{sec:results}
We finally highlight designs obtained for multiple rates and blocklengths to demonstrate the performance of the algorithms. In all but one of the code constructions, our initial frozen set comes from DEGA. In the case of the \mbox{$[128,42,L=32]$} code, our initial information set comes from RM-Polar. These initial sets are denoted $\mathcal{A}_0$. Throughout, the default precoder, $T_0$, comes from the time-invariant convolutional code with generator \mbox{$g_0 = [1,0,1,1,0,1,1]$.} The performance of all codes is plotted alongside the Normal Approximation (NA) \cite{NormalApx}, a finite-length approximation of capacity.

\begin{figure*}[t]
    \centering

    % Top-left
    \begin{minipage}[t]{0.48\textwidth}
        \centering
        \includegraphics[width=\linewidth]{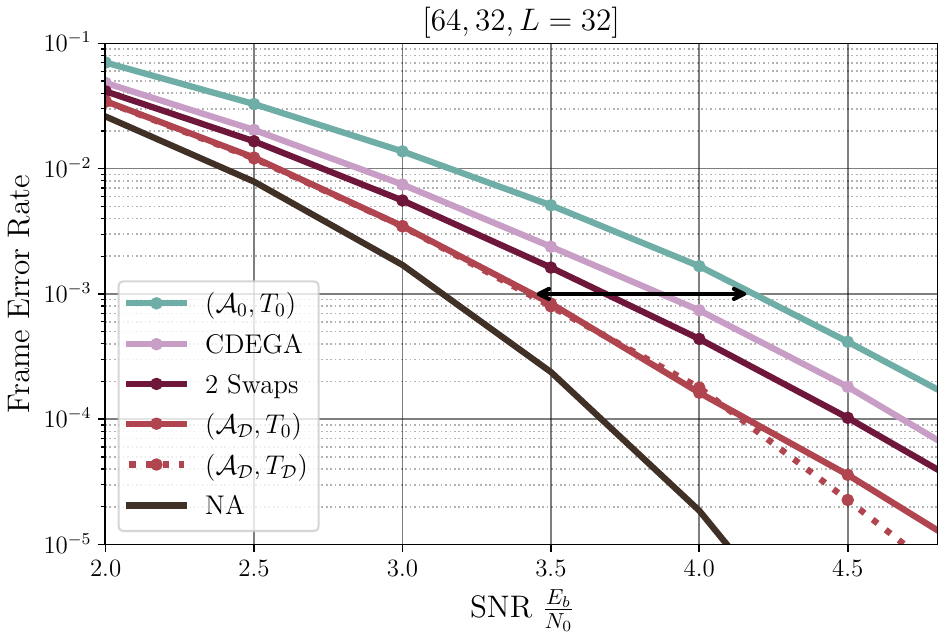}
        \captionof{figure}{$N = 64, K = 32, L =32$. 2 Swaps is the performance of Algorithm \ref{alg:set_design} after 2 iterations, evaluated with $T_0$.}
        \label{fig:64_32_FER}
    \end{minipage}
    \hfill
    % Top-right
    \begin{minipage}[t]{0.48\textwidth}
        \centering
        \includegraphics[width=\linewidth]{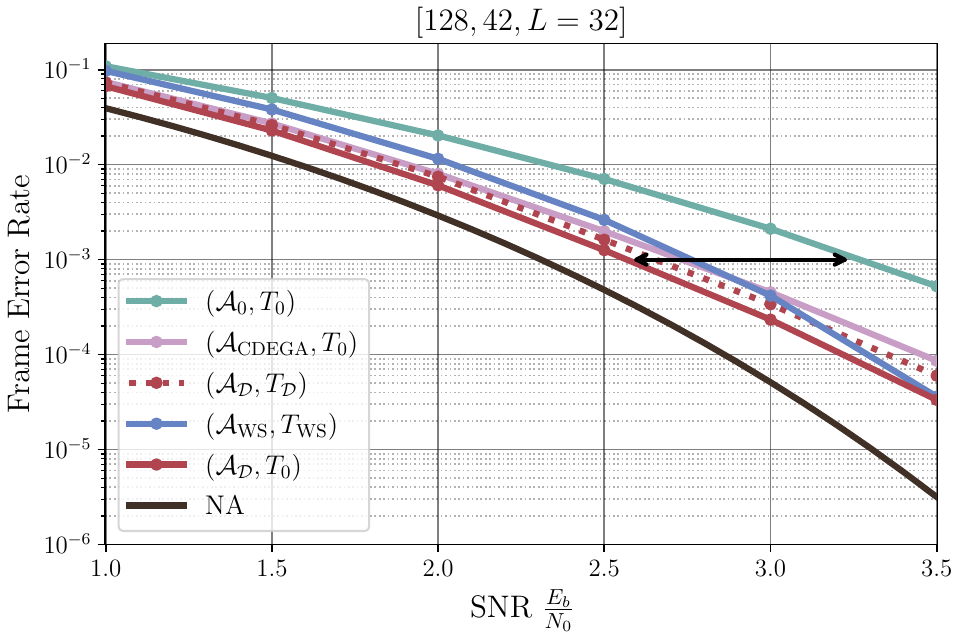}
        \captionof{figure}{$N = 128, K = 42, L =32$. WS uses the precoder from \cite{WeightedSum}.}
        \label{fig:128_42_FER}
    \end{minipage}

    \vspace{3mm}

    % Bottom-left
    \begin{minipage}[t]{0.48\textwidth}
        \centering
        \includegraphics[width=\linewidth]{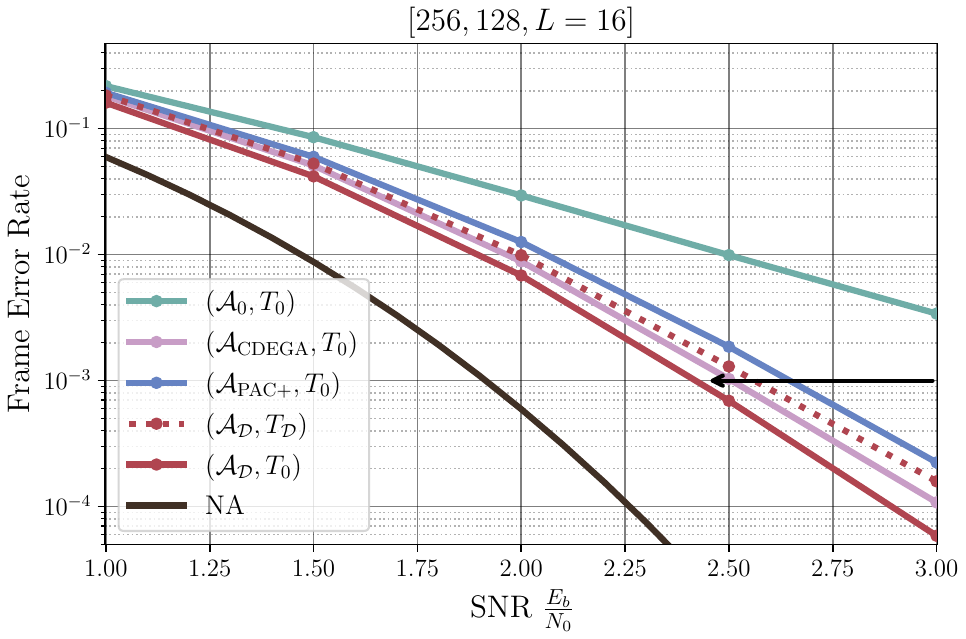}
        \captionof{figure}{$N = 256, K = 128, L = 16$.}
        \label{fig:256_128_FER}
    \end{minipage}
    \hfill
    % Bottom-right
    \begin{minipage}[t]{0.48\textwidth}
        \centering
        \includegraphics[width=\linewidth]{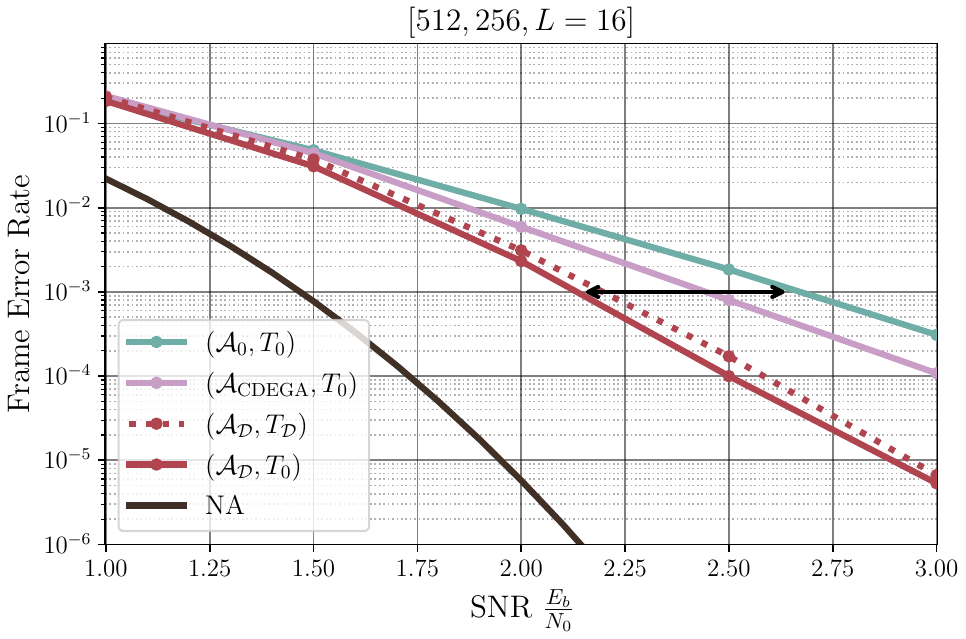}
        \captionof{figure}{$N = 512, K = 256, L = 16.$}
        \label{fig:512_256_FER}
    \end{minipage}
\vspace{-1mm}
\end{figure*}

The choice of the design SNR influences the outputs of our algorithms for two reasons: the construction $\mathcal{A}_0$ varies depending on SNR; but more importantly, the Monte Carlo simulations within the algorithms depend heavily on the simulation SNR. In general, we find that different design SNRs lead to different sets, which in turn leads to different final evaluations. As seen in Figures \ref{fig:64_32_FER}-\ref{fig:512_256_FER}, the performance of different constructions is not easily distinguishable at low SNRs, and the relevant metrics for Algorithms \ref{alg:algorithmCDEGA}-\ref{alg:precoder_design} tend to cluster. At high design SNR, the performance and metrics tend to disperse, and the algorithms are much more likely to make a near-optimal decision at each iteration. However, at too high of an SNR, errors may be too infrequent, leading to large variance and inaccuracies within the Monte Carlo simulation. 

Highlighting the gain at a target \ac{FER} of $10^{-3}$, all codes, as seen in Figures \ref{fig:64_32_FER}-\ref{fig:512_256_FER}, improve over $\mathcal{A}_0$, with gains of over .5 dB, and Figures \ref{fig:128_42_FER}-\ref{fig:256_128_FER} have gain approaching .75 dB. Additionally, we compare against two highly effective set constructors from the literature, weighted sum (WS) for $N=128$ \cite{WeightedSum} and PAC+ for $N=256$ \cite{rowshanFormationMinweightCodewords2023}. Here, we are able to find a slight gain in performance. The performance of the \ac{CDEGA} construction in Algorithm \ref{alg:algorithmCDEGA} is also shown. In all cases, the CDEGA set outperforms $\mathcal{A}_0$, however, it is less consistent than Algorithm \ref{alg:set_design}, as $N=64$ and $N=512$ have much smaller gains than at $N=128$ and $N=256$.

Focusing only on the information sets constructed, Algorithm \ref{alg:set_design} does not swap that many bits, exchanging as few as $4$ and as many as $7$ when $N = 512$. \mbox{Figure \ref{fig:64_32_FER}} plots the performance after just $2$ swaps, showing Algorithm \ref{alg:set_design} finds large gains within only a few iterations. The addition and removal of information bits, as well as other design parameters, can be seen in Table \ref{tab:sets}. Additionally, the information bits removed are typically much later in the sequential decoding process than the bits added. This more evenly spreads frozen bits throughout the information set, allowing more opportunities for a precoder to provide benefit.

Furthermore, as seen in Figures \ref{fig:64_32_FER}-\ref{fig:512_256_FER}, we are able to obtain similar FER with custom precoders that have far fewer off-diagonal elements. As seen in Table \ref{tab:prec}, two of the precoders are designed with $20$ off-diagonal elements, with $N = 256$ and $N = 512$ having $16$ and $26$ elements, respectively. Interestingly, $N = 256$ has the smallest number of off-diagonal elements, yet also has the largest degradation in performance when compared to $T_0$, possibly showing that precoders may lose benefits when they are too sparse. 

We do not claim that these sparse precoders have lower complexity than the generic precoder $T_0$. In fact, simple convolutional algorithms can certainly outperform sparse, time-varying matrices in most cases. These matrices exist as a way to isolate the benefits of \ac{PAC} and further analyze how the precoder interacts with the decoder. With few enough elements, it becomes possible to analyze which off-diagonal elements were chosen to give performance similar to $T_0$. Table \ref{tab:prec} shows which bits are both the \textit{source bits}, meaning information bits that are the input to the precoder, and which bits are the \textit{target bits}, meaning frozen bits that are the output of the precoder. Within the source bits, we highlight new bits that were added to $\calA_\calD$ throughout Algorithm \ref{alg:set_design}. Likewise, within the target bits, we highlight bits that were removed from the initial information set from Algorithm \ref{alg:set_design}.  Many of the bits newly added information bits show up as source bits, showcasing that their addition may be beneficial only in the presence of a precoder. Furthermore, in all cases except $N=256$, every information bit that was removed is used as a precoder target bit (even in the $N = 256$ exception, three of the four are used). This highlights the fact that despite being removed from $\mathcal{A}$, these bits still have high reliability (as evidence by their inclusion in traditional sets) and therefore the best performance comes from using them as the target of a precoder. This can be validated through CDEGA analysis, which often show that the precoder target bits have very high posteriors after an error occurs on the source bits. As stated in previous sections, the more a target bit is selected by source bits, it is less likely that the parity of the target bit is correct, and the benefit of the target bit may not be realized. 

Lastly, Table \ref{tab:dmin} shows the minimum Hamming distance, $d_\textrm{min}$ as well $A_{d_\textrm{min}}$, the number of codewords with distance $d_\textrm{min}$. In all cases, all algorithms improve the distance spectrum. There are a few aspects we highlight. In all cases except $N = 64$, the time-varying $T_\calD$ has a larger $A_{d_\textrm{min}}$ than $T_0$. This is consistent with the driving mechanism of $T_\calD$ not being the weight. $T_0$, which is denser than $T_\calD$, is more apt for randomization and improving the spectrum. However, despite differences in spectrum, all three constructions perform similarly for $N=128,256$. 

\section{Conclusion}
\label{sec:conc}
We present methods of analysis and construction of \acl{PAC} codes tailored towards \acl{SCL} decoding at list sizes too small to approximate \ac{ML} decoding. These methods are primarily focus on tracking the \acl{PM}, the underlying decision-making metric within the decoder, as well as analyzing the bits that contribute the most to the errors. We do not focus on weight spectrum, as metrics more focused on polarization and the decoder can be more important than weight in regimes where the list size is moderate. Using the two methods presented, \acl{CDEGA} and dangerous-path spectrum, we construct codes that improve upon state-of-the-art, as well as precoders that remain competitive when extremely sparse.

\newcolumntype{C}{>{\centering\arraybackslash}X}

\begin{table*}[t] 
\caption{Set Designs}
\label{tab:sets}
\centering 
\begin{tabularx}{\textwidth}{ |>{\hsize=1.0\hsize}C |>{\hsize=0.8\hsize}C |>{\hsize=0.8\hsize}C |>{\hsize=1.2\hsize}C |>{\hsize=1.2\hsize}C |}
 
\hline
Code Parameters &
Design SNR &
Initial Set &
Information Bits Added &
Information Bits Removed \\
\hline
$[64,32,\ell=32]$ & $3.5$ & DEGA & $13, 14, 21, 22$ &  $41, 49, 52, 53$  \\ 
\hline
$[128,42,\ell=32]$ & $3$ & RM-Polar & $ 27, 51, 54, 77, 78$ & $111, 116, 117, 123, 124$  \\ 
\hline
$[256,128,\ell=16]$ & $2$ & DEGA  & $31, 47, 86, 142$ & $208, 224, 233, 243$\\ 
\hline
$[512,256,\ell=16]$ & $2.5$ & DEGA & $63, 117, 121, 171, 216, 228, 337$ & $419, 465, 476, 479, 496, 501, 506$ \\ 
\hline
\end{tabularx} 
\end{table*}

\begin{table*}[t] 
\caption{Precoder Designs}
\label{tab:prec}
\centering 
\begin{tabularx}{\textwidth}{|>{\hsize=.775\hsize}C |>{\hsize=.1\hsize}C |>{\hsize=2.125\hsize}C |>{\hsize=1\hsize}C |}

\hline
Code Parameters &
$|T|$  &
Precoder Source Bits &
Precoder Target Bits \\
\hline
$[64,32,\ell=32]$ & $20$  & $\highlight{13}, \highlight{14}, \highlight{21}, \highlight{22}, 26, 28, 29, 38, 42, 44, 50$ & $25, 35, 37, \highlightb{41}, \highlightb{49}, \highlightb{52}, \highlightb{53}$\\ 
\hline
$[128,42,\ell=32]$ & $20$  & \makecell{$\highlight{51}, \highlight{54}, \highlight{77}, \highlight{78}, 86, 89, 90, 92, 105, $\\$ 106, 108, 109, 114, 120, 121, 122$} & $83, 99, \highlightb{111}, \highlightb{116}, \highlightb{117}, \highlightb{123}, \highlightb{124}$ \\ 
\hline
$[256,128,\ell=16]$ & $16$  & \makecell{$172, 178, 186, 202, 204, 212,$\\$ 216, 226, 228, 232, 240, 242$} & $\highlightb{224}, \highlightb{233}, \highlightb{243}$ \\ 
\hline
$[512,256,\ell=16]$ & $26$  & \makecell{$\highlight{228}, 313, \highlight{337}, 340, 344, 348, 353, 356, 360, 368, 391, 401, $\\$  406, 408, 417, 418, 420, 424, 450, 452, 456, 458, 480, 481$}& \makecell{$271, 323, 386, 387, 388, 389, 416, $\\$ \highlightb{419}, \highlightb{465}, \highlightb{476}, \highlightb{479}, \highlightb{496}, \highlightb{501}, \highlightb{506}$}\\ 
\hline
\end{tabularx} 
\end{table*}

\begin{table*}[t] \centering \caption{Minimum distance and multiplicity for different codes.} \label{tab:dmin} \renewcommand{\arraystretch}{1.1} \begin{tabular}{|c|c|c|c|c|c|c|c|c|c|} \hline \multirow{3}{*}{$N$} & \multirow{3}{*}{$K$} & \multicolumn{8}{c|}{Construction} \\ \cline{3-10} && \multicolumn{2}{c|}{$(\mathcal{A}_0, T_0)$} & \multicolumn{2}{c|}{$(\mathcal{A}_{\textrm{CDEGA}}, T_0)$} & \multicolumn{2}{c|}{$(\mathcal{A}_\calD, T_\calD)$} & \multicolumn{2}{c|}{$(\mathcal{A}_\calD, T_0)$} \\ 

\cline{3-10} && $d_{\min}$ & $A_{d_{\min}}$ & $d_{\min}$ & $A_{d_{\min}}$ & $d_{\min}$ & $A_{d_{\min}}$ & $d_{\min}$ & $A_{d_{\min}}$ \\ \hline

64 & 32 & 8 & 472 & 8 & 160 & 8 & 12  & 8 & 15 \\ \hline
128 & 42 & 16 & 792 & 16 & 94 & 16 & 58 & 16 & 21 \\ \hline 
256 & 128 & 8 & 96 & 16 & 7104 & 16 & 14168 & 16 & 3340  \\ \hline
512 & 256 & 16 & 36256 & 16 & 14832 & 16 & 1352 & 16 & 374 \\ \hline 

\end{tabular} \end{table*}

% ============================================================
% Stage 1
% ============================================================
\begin{algorithm*}[t]
\caption{Information-Set Design}
\label{alg:set_design}
\DontPrintSemicolon

\KwIn{
$N$, $K$, $L$, design SNR,
initial precoder $T_0$, swap sizes $\mathcal{S}$,
number of iterations $I_{\max}$
}
\KwOut{Optimized information set $\mathcal{A}^{\star}$}

Compute bit-channel reliability vector $\mathbf{r}$\;
Initialize $\mathcal{A}$ using the $K$ most reliable positions\;

Simulate SCL decoding to compute the frame error rate $Z$, and dangerous-path spectrum $\calD$
\[
    Z,\mathcal{D}
    \gets
    \operatorname{FullMC}(\mathcal{A},T_0).
\]

\For{$\ell \gets 1$ \KwTo $I_{\max}$}{

    Compute dangerous-path statistics for all bit positions\;

    Rank information positions $i\in\mathcal{A}$ by their
    removal score $R_i$ (chosen to be $\delta_i$)\;

    Rank frozen positions $j\in\mathcal{A}^{c}$ by their
    add reliability score $Q_j$ (chosen to be $q_{0.1}(|\lambda_j|)$)\;

    Form a small removal pool
    $\mathcal{R}\subseteq\mathcal{A}$
    from the highest-ranked removal positions\;

    Form a small addition pool
    $\mathcal{P}\subseteq\mathcal{A}^{c}$
    from the highest-ranked addition positions\;

    $\mathcal{C}\gets\varnothing$\;

    \ForEach{$s\in\mathcal{S}$}{
        \ForEach{$R\subseteq\mathcal{R}$ with $|R|=s$}{
            \ForEach{$P\subseteq\mathcal{P}$ with $|P|=s$}{
                Form the candidate information set
                \[
                    \widetilde{\mathcal{A}}
                    =
                    (\mathcal{A}\setminus R)\cup P;
                \]

                Add $\widetilde{\mathcal{A}}$ to $\mathcal{C}$\;
            }
        }
    }

    Rank the candidate information sets in $\mathcal{C}$ and retain
    the highest-ranked candidates\;

    \ForEach{$\widetilde{\mathcal{A}}\in\mathcal{C}$}{
        Estimate
        \[
            \widetilde{Z}_{\mathrm{small}}
            (\widetilde{\mathcal{A}})
            =
            \operatorname{MiniMC}
            (\widetilde{\mathcal{A}},T_0)
        \]
        using a reduced Monte-Carlo budget\;
    }

    Select the $M$ candidates having the smallest
    $\widetilde{Z}_{\mathrm{small}}
    (\widetilde{\mathcal{A}})$\;

    \ForEach{selected candidate $\widetilde{\mathcal{A}}$}{
        Re-evaluate using the full Monte-Carlo budget:
        \[
            \widetilde{Z}_{\mathrm{full}}
            (\widetilde{\mathcal{A}}),
            \widetilde{\mathcal{D}}
            (\widetilde{\mathcal{A}})
            \gets
            \operatorname{FullMC}
            (\widetilde{\mathcal{A}},T_0).
        \]
    }

    Let
    \[
        \widehat{\mathcal{A}}
        =
        \arg\min_{\widetilde{\mathcal{A}}}
        \widetilde{Z}_{\mathrm{full}}
        (\widetilde{\mathcal{A}}).
    \]

    \eIf{
        $\widetilde{Z}_{\mathrm{full}}
        (\widehat{\mathcal{A}})<Z$
    }{
        $\mathcal{A}\gets\widehat{\mathcal{A}}$\;
        $Z\gets
        \widetilde{Z}_{\mathrm{full}}
        (\widehat{\mathcal{A}})$\;
        $\mathcal{D}\gets
        \widetilde{\mathcal{D}}
        (\widehat{\mathcal{A}})$\;
    }{
        \textbf{break}\;
    }
}

$\mathcal{A}^{\star}\gets\mathcal{A}$\;

\Return{$\mathcal{A}^{\star}$}\;
\end{algorithm*}

% ============================================================
% Stage 2
% ============================================================
\begin{algorithm*}[t]
\caption{Sparse PAC Precoder Construction}
\label{alg:precoder_design}
\DontPrintSemicolon

\KwIn{
Information set $\mathcal{A}^{\star}$,
path-loss maximum edge span $b$,
edge-set size $k$,
maximum number of iterations $E_{\max}$,
replay threshold $\tau$
}
\KwOut{Optimized sparse precoder $T^{\star}$}

Initialize $T\gets I_N$\;

Simulate SCL decoding to compute the frame error rate $Z$, dangerous-path spectrum $\calD$, and replay trials $\calR$
\[
    Z,\mathcal{D},\mathcal{R}
    \gets
    \operatorname{FullMC}
    (\mathcal{A}^{\star},T,\tau),
\]

\For{$\ell\gets1$ \KwTo $E_{\max}$}{

    $\mathcal{E}\gets\varnothing$\;

   \ForEach{$i\in\mathcal{A}^{\star}$}{
        \ForEach{frozen bit $j\in\mathcal{F}$
        with $i<j\leq N - 1$ }{
            \If{$T_{ij}=0$}{
                Compute the dangerous-path rescue score $S(i,j)$ from $\mathcal{D}$\;

                \If{$S(i,j)>0$}{
                    Add $(i,j)$ to $\mathcal{E}$\;
                }
            }
        }
    }

    Rank the candidate edges in $\mathcal{E}$ by $S(i,j)$ and
    retain a small pool $\mathcal{E}_{\mathrm{top}}$\;

    $\mathcal{C}\gets\varnothing$\;

    \ForEach{$E\subseteq\mathcal{E}_{\mathrm{top}}$ with $|E|=k$}{
        Compute the joint rescue score $S(E)$ over the current
        dangerous-path population\;

        Add $E$ to $\mathcal{C}$\;
    }

    Rank the edge sets $E\in\mathcal{C}$ by $S(E)$ and retain
    the highest-ranked candidate sets\;

    \ForEach{$E\in\mathcal{C}$}{
        Form the candidate precoder
        \[
            T_E \gets T + E ;
        \]

        Replay the stored error and near-miss realizations under $T_E$:
        \[
            Z_{\mathrm{rep}}(E)
            \gets
            \operatorname{Replay}
            (\mathcal{A}^{\star},T_E,\mathcal{R}) .
        \]
    }

    Select the $M$ candidates having the smallest
    $Z_{\mathrm{rep}}(E)$\;

    \ForEach{selected edge set $E$}{
        Re-evaluate using the full Monte-Carlo budget:
        \[
            \widetilde Z_{\mathrm{full}}(E),
            \widetilde{\mathcal{D}}(E),
            \widetilde{\mathcal{R}}(E)
            \gets
            \operatorname{FullMC}
            (\mathcal{A}^{\star},T_E,\tau).
        \]
    }

    Let
    \[
        \widehat E
        =
        \arg\min_E
        \widetilde Z_{\mathrm{full}}(E).
    \]

    \eIf{$\widetilde Z_{\mathrm{full}}(\widehat E)<Z$}{
        $T\gets T_{\widehat E}$\;
        $Z\gets\widetilde Z_{\mathrm{full}}(\widehat E)$\;
        $\mathcal{D}\gets\widetilde{\mathcal{D}}(\widehat E)$\;
        $\mathcal{R}\gets\widetilde{\mathcal{R}}(\widehat E)$\;
    }{
        \textbf{break}\;
    }
}

$T^{\star}\gets T$\;

\Return{$T^{\star}$}\;
\end{algorithm*}

\clearpage
\clearpage
\IEEEtriggeratref{14}
\bibliographystyle{IEEEtran}
\bibliography{references.bib}

\clearpage

\end{document}